\documentclass[a4paper,fleqn,usenatbib]{mnras}

\usepackage[T1]{fontenc}
\usepackage{ae,aecompl}

\usepackage{graphicx}	
\usepackage{amsmath}	
\usepackage{amssymb}	
\usepackage{url}
\usepackage{gensymb}
\usepackage{subcaption}
\usepackage{float}
\usepackage{placeins}

\usepackage{tikz}
\usepackage{graphicx}
\usepackage{ulem}
\usepackage{tabularx} 

\usepackage[font=small,skip=0pt]{caption}
\usepackage [english]{babel}
\usepackage [autostyle, english = american]{csquotes}
\MakeOuterQuote{"}

\title[VST ATLAS Galaxy Cluster Catalogue I]{VST ATLAS Galaxy Cluster Catalogue I: cluster detection and mass calibration}

\author[B. Ansarinejad]{
Behzad Ansarinejad\thanks{E-mail: behzad.ansarinejad@unimelb.edu.au}$^{1,2}$, David Murphy$^{3}$, Tom Shanks$^{1}$, Nigel Metcalfe$^{1}$
\\
$^{1}$Centre for Extragalactic Astronomy, Department of Physics, Durham University, South Road, Durham DH1 3LE, UK\\
$^2$School of Physics, University of Melbourne, Parkville, VIC 3010, Australia\\
$^{3}$Institute of Astronomy, University of Cambridge, Madingley Road, Cambridge CB3 0HA
}

\date{Accepted XXX. Received YYY; in original form ZZZ}

\pubyear{2022}
\begin{document}
\label{firstpage}
\pagerange{\pageref{firstpage}--\pageref{lastpage}}
\maketitle

\begin{abstract}
Taking advantage of $\sim4700$ deg$^2$ optical coverage of the Southern sky offered by the VST ATLAS survey, we construct a new catalogue of photometrically selected galaxy groups and clusters using the {\sc orca} cluster detection algorithm. The catalogue contains $\sim 22,000$ detections with $N_{200}>10$ and $\sim9,000$ with $N_{200}>20$. We estimate the photometric redshifts of the clusters using machine learning and find the redshift distribution of the sample to extend to $z\sim0.7$, peaking at $z\sim0.25$. We calibrate the ATLAS cluster mass-richness scaling relation using masses from the MCXC, Planck, ACT DR5 and SDSS redMaPPer cluster samples. We estimate the ATLAS sample to be $>95\%$ complete and $>85\%$ pure at $z<0.35$ and in the $M_{\textup{200m}}$>$1\times10^{14}h^{-1}$\(\textup{M}_\odot\) mass range. At $z<0.35$, we also find the ATLAS sample to be more complete than redMaPPer, recovering a $\sim40\%$ higher fraction of Abell clusters. This higher sample completeness places the amplitude of the $z<0.35$ ATLAS cluster mass function closer to the predictions of a $\Lambda$CDM model with parameters based on the Planck CMB analyses, compared to the mass functions of the other cluster samples. However, strong tensions between the observed ATLAS mass functions and models remain. We shall present a detailed cosmological analysis of the ATLAS cluster mass functions in paper II. In the future, optical counterparts to X-ray-detected eROSITA clusters can be identified using the ATLAS sample. The catalogue is also well suited for auxiliary spectroscopic target selection in 4MOST. The ATLAS cluster catalogue is publicly available at \url{http://astro.dur.ac.uk/cosmology/vstatlas/cluster_catalogue/}.
\end{abstract}




\section{Introduction}
\label{sec:introduction}

In the context of hierarchical structure formation, primordial peaks in the density field of the early Universe collapse and merge, leading to the incremental formation of gravitationally bound halos of increasing mass (e.g. \citealt{Peebles1980}). Galaxy clusters are the largest of these gravitationally bound structures in the Universe, occupying the extreme tail of the halo mass function. As a result, the evolution of galaxy cluster abundance with mass and redshift is extremely sensitive to variations in cosmological parameters. Precise observations of large numbers of clusters over a wide range of masses and redshifts can therefore provide powerful cosmological constraints (see reviews by \citealt{Allen2011}; \citealt{Weinberg2013}). In the context of the $\Lambda$CDM model, the study of galaxy clusters can place independent constraints on $\Omega_\textup{m}$, the matter density parameter (\citealt{Evrard1997}; \citealt{Schuecker2003}), $\Omega_{\Lambda}$, the dark energy density parameter and $\omega$, the dark energy equation of state parameter (\citealt{Morandi2016}), as well as $\sigma_8$, the normalization of the matter power spectrum on the scale of $8\ h^{-1}$Mpc (e.g. \citealt{White1993}). Furthermore, extensions to $\Lambda$CDM such as massive neutrinos can be investigated via the study of galaxy cluster number counts (\citealt{Costanzi2013}; \citealt{Roncarelli2015}), while non-standard modified gravity models can be constrained via the study of cluster properties (\citealt{Brownstein2009}; \citealt{Llinares2013}; \citealt{Planck2016XIII}; \citealt{Bocquet2019}). Galaxy clusters and groups also provide useful tools in studying galaxy evolution in extreme environments and their physical properties could aid our understanding of structure formation, providing vital information on collapse of dark matter and evolution of baryons in dark matter potentials (see reviews by \citealt{Rosati2002}; \citealt{Voit2005} and \citealt{Kravtsov2012}). 

In the optical regime, the Sloan Digital Sky Survey (SDSS; \citealt{York2000}) was one of the first wide-field surveys covering $\sim10,000$ deg${^2}$ of the sky, which formed the basis of the MAXBCG \citep{Koester2007} and redMaPPer \citep{Rykoff2014} cluster catalogues. The VST ATLAS survey \citep{Shanks2015}, which is the basis of this work, later provided coverage of $\sim4700$ deg${^2}$ over the southern sky to similar depths as SDSS, but with superior seeing. Other notable recent and upcoming photometric surveys include Dark Energy Survey (DES; \citealt{DES2005}; \citealt{Rykoff2016}), Pan-STARRS (\citealt{Kaiser2002}; \citealt{Ebeling2013}), the Kilo-Degree Survey (KiDS; \citealt{deJong2013}; \citealt{Radovich2017}), the Hyper-Suprime Camera (HSC; \citealt{Aihara2017}), and the Legacy Survey of Space and Time (LSST; \citealt{LSST2012}), 4-metre Multi-Object Spectroscopic Telescope (4MOST; \citealt{deJong2019}) and Euclid \citep{Laureijs2011}, providing enormous quantities of data upon which the search for clusters of galaxies has and will be based.
 
In the absence of spectroscopic redshifts, galaxy redshifts can be estimated photometrically based on techniques such as SED template fitting (e.g. Hyperz; \citealt{Bolzonella2000}), using Bayesian probabilistic methods (e.g. BPZ; \citealt{BPZ2011}) or via machine learning approaches utilising artificial neural networks or boosted decision trees (e.g. ANNz2; \citealt{Sadeh2016}). Due to their large uncertainties, however, photometric redshifts alone are not sufficient for accurate identification of galaxy clusters via 3D reconstruction of the distribution of galaxies. A commonly adopted method of detecting galaxy clusters in the optical regime is taking advantage of the cluster \textit{red sequence} (\citealt{Baum1959}; \citealt{Bower1992}; \citealt{Gladders2000}). The red sequence refers to the tight correlation followed by the cluster galaxies in the colour-magnitude space, this feature arises due to galaxy clusters mostly containing early-type elliptical and lenticular galaxies which consist of passively evolving stellar populations giving rise to strong metal absorption lines at wavelengths blue-wards of $4000$\AA. As a result, the majority of cluster members appear red in colour and occupy a narrow ridge in the colour-magnitude space, when viewed through broadband photometric filters that straddle the $4000$\AA\ break. In this way cluster galaxies can be isolated from active, star-forming field galaxies.  Algorithms utilising the red sequence for cluster detection include \cite{Koester2007}; \cite{Gladders2007}; \cite{Thanjavur2009}; \cite{Hao2010}; \cite{Rykoff2014} as well as The Overdense Red-sequence Cluster Algorithm ({\sc orca}; \citealt{Murphy2012}) used in this work.

Galaxy clusters have also been detected in X-ray by exploiting the radiation due to thermal bremsstrahlung and line emission from the Inter Cluster Medium (\citealt{Cavaliere1976}; \citealt{Allen2002}). In this work, we compare our optically detected cluster catalogue with the `MCXC: a meta-catalogue of X-ray detected clusters of galaxies' (\citealt{Piffaretti2011}) which contains $\sim1700$ X-ray detected clusters. In the near future, an ongoing all-sky X-ray survey eROSITA \citep{Merloni2012} with the aim of detecting $\sim100,000$ galaxy clusters out to $z>1$ will significantly increase the cluster sample size in the X-ray regime. The cluster catalogue introduced in this work will provide a valuable resource which could be used to better characterise the selection function of the eROSITA cluster in terms of sample completeness and purity.

Inverse Compton scattering due to interaction with high energy electrons in the ICM provides a boost in energy to the cosmic microwave background (CMB) photons passing through clusters, making them detectable via a phenomenon known as the Sunyaev-Zel'dovich (SZ) effect (\citealt{Sunyaev1980}). With the benefit of a detection signature that is essentially redshift independent, the SZ effect offers a new window on the cluster population providing a nearly mass-limited sample at high redshifts, where cluster abundance can place sensitive constraints on cosmological parameters (\citealt{Carlstrom2002}; \citealt{Planck_Collaboration2014}). In recent years, several SZ surveys have been undertaken with South Pole Telescope (SPT; \citealt{Carlstrom2011}) survey, the Atacama Cosmology Telescope (ACT; \citealt{Fowler2007}), and Planck \citep{Tauber2010}, providing the first SZ-selected cluster samples (\citealt{Vanderlinde2010};  \citealt{Menanteau2010}; \citealt{Planck2011}; \citealt{Reichardt2013}; \citealt{Hasselfield2013}). In this work, we also draw a comparison between our cluster catalogue and the second Planck catalogue of SZ sources (henceforth Planck SZ; \citealt{PSZ2_cat_2016}), which contains 1653 detections, out of which 1203 are confirmed clusters based on identification of counterparts in external datasets. Similarly, we compare the ATLAS cluster catalogue with SZ detections from the ACT DR5 sample \citep{AdvACT2020}, which has a large overlap with part of ATLAS in the Southern Galactic Cap (SGC). The ACT DR5 sample contains over 4000 cluster detections in a survey area of $13,168$ deg$^2$, probing lower cluster masses than the Planck SZ sample.

We then present a calibration of the ATLAS cluster mass-richness scaling relation using cluster masses from the MCXC, ACT DR5, Planck and SDSS redMaPPer samples. Our final aim in this work is to compare the observed mass function of the ATLAS cluster catalogue to theoretical predictions of $\Lambda$CDM based on the \cite{Tinker2008} model and other cluster samples. We choose the \cite{Tinker2008} model due to its common use in literature for cosmological analyses of various cluster samples over the past decade. (see e.g. \citealt{Vikhlinin2009}; \citealt{Allen2011}; \citealt{Bleem2015}; \citealt{Planck2016XXIV}; \citealt{Bocquet2019}; \citealt{DES2020}). We perform this comparison in five redshift bins covering the range $0.05<z<0.55$ in order to examine the evolution of the cluster mass functions.

The layout of this paper is as follows: In Section~\ref{sec:Dataset} we describe the VST ATLAS galaxy input catalogue. Then in Section~\ref{sec:Methodology} we present a brief description of the {\sc orca} algorithm, as well as a description of our machine learning approach to obtaining photometric redshifts for our clusters. We also present a description of our cluster richness and mass estimation. In Section~\ref{sec:Results}, we present the ATLAS cluster catalogue and compare our results with external multi-wavelength cluster samples, providing estimates of the completeness and purity of the ATLAS cluster catalogue. This is followed by a comparison of the observed cluster mass functions of the ATLAS, redMaPPer, Planck and ACT DR5 samples, versus theoretical predictions. Finally, we conclude by summarizing the cosmological implications in Section~\ref{sec:Conclusions}. Unless otherwise specified, we assume a \cite{Planck2016XIII} $\Lambda$CDM cosmology with $H_0=67.74$, $\Omega_{\Lambda}=0.6911$, $\Omega_m=0.3089$, $\sigma_8=0.8159$, present all magnitudes in the AB system and define our cluster masses as $M_{200\textup{m}}$ which is the mass enclosed in a halo with a density $200\times$ the mean matter density of the Universe.

\section{Dataset}
\label{sec:Dataset}

The VST ATLAS \citep{Shanks2015} is an European Southern Observatory (ESO) public survey of the southern sky, designed to provide optical imaging in $ugriz$ bands to similar depths as SDSS in the north. The data is taken using the Very Large Telescope Survey Telescope (VST; \citealt{Schipani2012}) a 2.61-m telescope with a $1\deg^2$ field of view located at the Paranal observatory. The total survey area consists of $4711 \deg^2$, with $2087 \deg^2$ in the Northern Galactic Cap (NGC) and $2624 \deg^2$ in the Southern Galactic Cap (SGC). The ATLAS coverage area is overlapped by the DES \citep{DES2005}, DESI \citep{DESI2016}, KiDS \citep{deJong2013} and eROSITA \citep{Merloni2012} surveys. In the future, the 4MOST \citep{deJong2019}, Euclid \citep{Laureijs2011} and LSST \citep{LSST2012} surveys will also provide multi-wavelength imaging and spectroscopic coverage of the VST ATLAS survey area.

Based on the ATLAS DR4 data, we produce a band-merged $griz$ catalogue using a pipeline developed with the `Starlink Tables Infrastructure Library Tool Set' ({\sc STILTS}; \citealt{Taylor2006}) framework. During band-merging, we include all objects with detections in a minimum of two bands. In this work, we utilise the VST ATLAS aperture magnitudes provided by the Cambridge Astronomical Survey Unit (CASU)\footnote{\url{http://casu.ast.cam.ac.uk/surveys-projects/vst/technical/catalogue-generation}} for obtaining galaxy colours while relying on the CASU Kron pseudo-total magnitude for imposing magnitude limits on our galaxy samples. The ATLAS Kron magnitudes are measured using circular apertures of $2\times$ the Kron radius, with the definition of the Kron radius given by \cite{Bertin1996}. The use of aperture magnitudes to measure galaxy colours was motivated by the inspection of the red-sequence of rich, spectroscopically confirmed Abell clusters, as well as comparison of galaxy colours between ATLAS Kron and aperture magnitudes and SDSS model magnitudes in a $\sim200$ deg$^2$ overlap area between the two surveys. These tests showed that aperture magnitudes have a lower level scatter than Kron around the red-sequence and when compared to SDSS. On the other hand comparison between ATLAS and SDSS shows that Kron magnitudes provide a more reliable measure of the galaxies' total magnitudes compared to aperture magnitudes, which could miss a larger fraction of the galaxy flux (see also \citealt{Shanks2015}).  We denote aperture magnitudes corresponding to the ATLAS \texttt{Aperture flux 5} using the subscript `A5'. This aperture has a radius of $2''$ and we apply the associated aperture correction labelled as \texttt{APCOR} in the CASU catalogue. For $g, r, i$ and $z$ bands, the mean values of \texttt{APCOR5} are 0.12, 0.12, 0.11 and 0.12 mags. Although these aperture corrections are derived for stars, they also provide a first-order seeing correction for faint galaxies. Where Kron pseudo-total magnitudes are used, we correct these to total magnitude for galaxies by applying a $-0.15$ mag offset. This value is chosen based on an empirical comparison of the ATLAS Kron and SDSS model magnitudes for galaxies in an overlap area between the two surveys as shown by \cite{Shanks2015}. We then correct all magnitudes for Galactic dust extinction $A_\textup{x}=C_\textup{x}E(B-V)$, with $x$ representing a filter ($griz$), taking the SDSS $C_\textup{x}$ values presented in \cite{Schneider2007} (3.793, 2.751, 2.086, and 1.479 for $griz$ respectively) and using the Planck $E(B-V)$ map (\citealt{Planck2014}).

In order to isolate galaxies from stars in the ATLAS data, we use the default morphological classifications supplied in CASU catalogues, a description of which can be found in \cite{Gonzalez2008}. Reflections from bright stars in the VST ATLAS data could lead to the formation of circular halos which in some cases can be misidentified as multiple extended sources and misclassified as galaxies by the CASU source detection algorithm. To overcome this, we mask circular regions around these bright stars based on cross-matching the input catalogue to the Tycho-2 Catalogue of the 2.5 million brightest stars \citep{Hog2000}. The Tycho bright stars are masked with radii varying according to their $V$-band magnitudes. For this purpose we choose the following radii based on visual inspection of stars with various magnitudes: $V<8$: $340''$; $8<V<9$: $80''$; $9<V<10$: $45''$; $10<V<11$: $30''$; $V>11$: $20''$. However, depending on the position of the star on the CCD chip, in some cases the halo can be off-centred from the stars. The remaining stellar halos and other major remaining artefacts such as nearby galaxies or satellite trails are manually removed by performing a visual inspection of the data.

The input catalogue given to {\sc orca} for cluster detection consists of objects detected in two adjacent bands (i.e. $g-r$, $r-i$ or $i-z$ which are used for cluster red-sequence detection). We also require objects to be classified as galaxies in a minimum of two bands. Here, we only require objects to be classified as galaxies in two bands as demanding a galaxy classification in all bands was deemed too strict, resulting in a $\sim10\%$ decrease in the overall sample size. The slight increase in stellar contamination introduced in this approach is unlikely to be a problem in our cluster detection, due to {\sc orca}'s reliance on the red sequence in selecting cluster members (see Section~\ref{sec:ORCA}). We do not include the $u$ band due to its shallow depth and the fact that the incompleteness of $u$-band observations at the time of conducting this work would result in large gaps across the survey area.

As the VST ATLAS survey observations are conducted in $17$ deg$^2$ blocks at constant Declination, taking data in a single band at a time, nightly variations in seeing, sky brightness and other observing conditions can result in slight variations in survey depth across different concatenations. After band merging, these slight fluctuations in object densities in some concatenations could result in artificial inhomogeneities across the sky. Consequently, we select magnitude limits of $g_{\rm Kron}<22.0$; $r_{\rm Kron}<21.6$; $i_{\rm Kron}<21.1$; $z_{\rm Kron}<19.9$ as a compromise between increasing the survey depth and increasing homogeneity between concatenations. The final input catalogues contain $\sim8,740,000$ galaxies in the SGC and $\sim7,825,000$ in the NGC of the survey.

\section{Methodology}
\label{sec:Methodology} 

\subsection{{\sc orca}: The cluster detection algorithm}
\label{sec:ORCA}

A detailed description of the `Overdense Red-sequence Cluster Algorithm' ({\sc orca}) which is used to create the ATLAS cluster catalogue can be found in \cite{Murphy2012}. The algorithm detects the red sequence in pairs of adjacent bands or colours (in our case $g-r$, $r-i$ or $i-z$). In the first stage, the algorithm applies a selection function to the input catalogue in the form of narrow slices in the colour-magnitude space. This photometric filtering separates galaxies within a specific redshift range from foreground and background objects, broadly isolating the cluster galaxies via the $4000$\AA\ break. By detecting clusters in two colours concurrently, foreground and background contamination can be significantly reduced, as galaxies follow unique tracks in different colour-redshift spaces. During this stage, the red sequence is isolated across a range of redshifts through modifications of the colour slice in successive runs of the algorithm, systematically scanning the entire photometric space.

Upon the application of photometric filtering, the algorithm estimates the surface density of the remaining galaxies by calculating the Voronoi diagram of their projected distribution on the sky. The Voronoi cells are then separated into overdense and underdense cells based on a user-specified probability threshold ($P_{\rm thresh}$), related to how likely they are to belong to a random distribution (for more details, see Section 3.4 of \citealt{Murphy2012}). Finally using the Friend-Of-Friends technique, the algorithm connects adjacent overdense cells until the density of the whole system falls below a user-defined critical density ($\sum_{\rm crit}$). At this stage, if the system has at least $N_{\rm gals}$ linked galaxies (in this case we set $N_{\rm gals}=5$), it is defined as a cluster.

We optimize the $P_{\rm thresh}$ and $\sum_{\rm crit}$ parameters for performance on VST ATLAS based on multiple runs of the {\sc orca} algorithm on a $\sim300$ deg$^2$ area and assessing its performance based on recovering the Abell, MCXC, redMaPPer and Planck SZ clusters in this region. We then set $\sum_{\rm crit}=2.5\bar{\sum}$ (where $\bar{\sum}$ is the mean galaxy density), and $P_{\rm thresh}=0.0125$, using the default for other adjustable parameters of the algorithm as these do not have a major effect on improving the results. As the adjustment to the colour slice is, by design, less than the width of the red sequence, the same cluster can be identified multiple times in successive runs of the algorithm.

\subsection{{\sc ANNz2}: Photometric redshift estimation}
\label{sec:ANNz2}

We make use of the publicly available {\sc ANNz2}\footnote{\url{https://github.com/IftachSadeh/ANNZ}} \citep{Sadeh2016} algorithm in order to obtain photometric redshift estimates for the VST ATLAS cluster members. For this purpose, we simultaneously make use of a combination of two machine learning methods offered by the algorithm; Artificial Neural Networks (ANNs) and Boosted Decision Trees (BDTs). After performing various tests, this approach was shown to produce the best RMS scatter in comparison to using ANNs or BDTs alone. Here the RMS error is given by:  

\begin{equation}
\sigma_{\Delta z/(1+z)}\equiv\sqrt{\frac{1}{n_{\rm gals}}\sum_{\rm gals}\left(\frac{z_{\rm photo}-z_{\rm spec}}{1+z_{\rm spec}}\right)^2},
\label{eq:RMS}
\end{equation}

where $n_{\rm gals}$ is the number of galaxies in the training set and $z_{\rm photo}$ and $z_{\rm spec}$ are estimated photometric and measured spectroscopic redshifts of these galaxies, respectively.

\begin{figure}
	\begin{subfigure}[t]{\columnwidth}
		\centering
		\includegraphics[width=0.9\columnwidth]{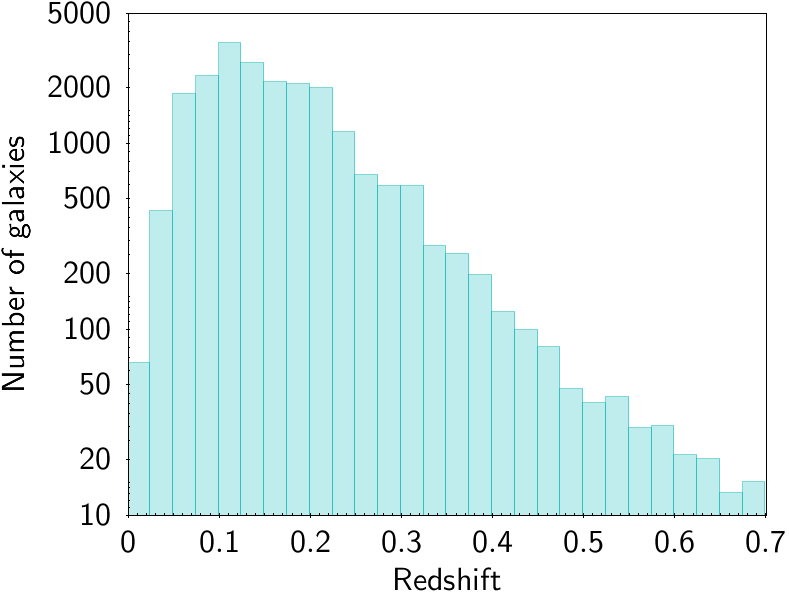}
	\end{subfigure}
	\caption[Redshift distribution of the ANNz2 machine learning training samples.]{The redshift distribution of the training sample used to estimate the photometric redshifts of the ATLAS galaxy cluster members by ANNz2.}
	\label{fig:spec_training_hist.png}
\end{figure}

To estimate photometric redshifts using machine learning, we are required to provide the algorithm with a training sample of galaxies with measured redshifts, which overlap the VST ATLAS survey area. For this purpose, we make use a total of 21,114 galaxies with redshifts obtained from various surveys as detailed in Table~\ref{tab:redshift_surveys}. In cases where the same objects have redshifts provided by more than one survey, we keep the redshift with the smaller uncertainty. Figure~\ref{fig:spec_training_hist.png} shows the redshift distribution of the galaxies included in the training set used in our photometric redshift estimation.

\begin{table*}
	\centering
	\caption[Details of the sub-samples used for training by the ANNz2 machine learning algorithm while estimating the VST ATLAS galaxy cluster photometric redshifts]{Details of the samples used in our photometric redshift training. Here the redshift coverage, the mean redshift of the samples, and the number of galaxies which corresponds to the number of objects with redshifts identified as ATLAS cluster members by the {\sc orca} algorithm.}
	\label{tab:redshift_surveys}
	\begin{tabular}{lcccc} 
		\hline
		 Sample & Redshift coverage & $\bar{z}$ & Number of galaxies & Reference \\
		\hline
		2dFGRS & $0.00<z<0.30$ & 0.13 & 9,426 & \cite{Colless2003} \\
		SDSS DR12 & $0.05<z<0.35$ & 0.15 & 5,260 & \cite{Alam2015} \\
		GAMA G23 & $0.05<z<0.50$ & 0.21 & 3,008 & \cite{Liske2015} \\
	    Primus & $0.05<z<0.60$ & 0.24 & 177 & \cite{Cool2013} \\
	    2dFLenS & $0.05<z<0.75$ & 0.26 & 2,717 & \cite{Wolf2017} \\
	    BOSS LOWZ & $0.05<z<0.45$ & 0.26 & 375 & \cite{Dawson2013} \\
	    BOSS CMASS & $0.40<z<0.75$ & 0.52 & 141 & \cite{Dawson2013} \\
		\hline
        \multicolumn{5}{p{\textwidth}}{}
	\end{tabular}
\end{table*}

In order to improve our photometric redshift training, in addition to the ATLAS $griz$ bands, we add the W1 and W2 magnitudes from the "unblurred coadds of the WISE imaging" catalogue (unWISE; \citealt{Schlafly2019})\footnote{We note that despite the $6''$ WISE PSF, thanks to the improved modelling of the blended sources in the unWISE catalogue, the addition of W1 and W2 information to our machine learning training still proved useful in improving our photo-z RMS scatter.}. Here we use a $2''$ radius to match ATLAS and unWISE sources and we correct for galactic dust extinction in W1 and W2 bands by subtracting $0.18\times E(B-V)$ and $0.16\times E(B-V)$ from the W1 and W2 magnitudes respectively. These $E(B-V)$ values are taken from the same Planck dust map \citep{Planck_Collaboration2014} used to correct the ATLAS magnitudes for galactic dust extinction and the $0.18$ and $0.16$ coefficients are taken from \cite{Yuan2013}.

Finally, we calculate the weighted mean photometric redshift of our clusters using: $\bar{z}=(\sum z_{\rm i}/{\sigma_{\rm i}}^2)/(\sum1/{\sigma_{\rm i}}^2)$, where $z_{\rm i}$ and $\sigma_{\rm i}$ are the photometric redshift and photometric redshift uncertainty of the $i$-th cluster member. The uncertainty on the cluster redshift is then given by the standard error on the mean, and we verify the uncertainty using the Jackknife technique.

\subsection{Cluster catalogue post-processing}
\label{sec:post_processing}

As discussed in 3.6 of \cite{Murphy2012}, multiple detections of the same cluster found in different colour-magnitude spaces are merged together by {\sc orca} based on five tests of `cluster similarity'. These criteria are based on the similarity of the clusters' red-sequence, the extent of spatial overlap and the number of common galaxies between the two detections.

In this work, we utilise our photometric redshifts to further merge overlapping cluster detections that are likely to belong to the same system based on the following criteria:

\begin{itemize}
\item \textbf{Spatial overlap:} A cluster centre lies within the cluster radius of a nearby cluster. Here, the cluster centre is defined as the mean RA and Dec of its cluster member and cluster radius is defined as the angular separation between the furthest cluster member and the cluster centre. In addition, we require at least one of the criteria below to be satisfied in order for overlapping detections to be considered as part of the same system.

\item \textbf{Photometric redshift overlap:} The error weighted mean cluster photometric redshift lies within one standard error of another spatially overlapping cluster.

\item \textbf{Red-sequence overlap:} The mean colour of a cluster's members (in either $g-r$, $r-i$ or $i-z$) is within one standard error of another spatially overlapping cluster's mean colour.

\end{itemize}

\begin{figure}
	\begin{subfigure}[hbt]{\columnwidth}
		\centering
		\includegraphics[width=\columnwidth]{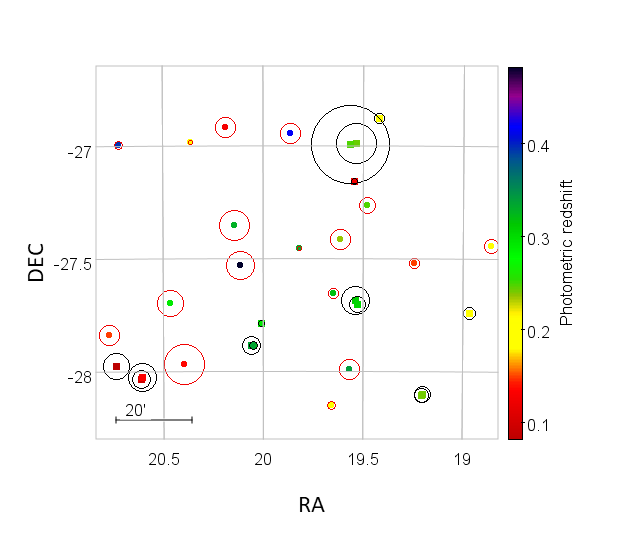}
		\label{fig:clusters_merged_radius_photoz_colour.png}
	\end{subfigure}
	\caption[Merging overlapping clusters]{The outcome of merging overlapping clusters based on spatial and photometric redshift, or red-sequence overlap. Here, circles show the corresponding cluster radii used to identify spatially overlapping clusters, with the black circles marking clusters that are merged.}
	\label{fig:merging}
\end{figure}

Requiring the combination of spatial overlap, with either photometric redshift or red-sequence overlap results in the merging of $9\%$ of the clusters in the catalogue. Figure~\ref{fig:merging} shows an example of merged overlapping clusters based on the above criteria.

\subsection{Cluster richness ($N_{200}$)}
\label{sec:Cluster_richness}

In order to provide an estimate of cluster richness, we first count the number of cluster galaxies within a $1 h^{-1}$ Mpc aperture of the cluster centre, $N_{\textup{1Mpc}}$, which in the $i$-band, are fainter than the Brightest Cluster Galaxy (BCG), and brighter than $0.4L_*$. Here $L_*$ is the characteristic luminosity in the Schechter luminosity function, and following \cite{Reyes2008}, we take the $z=0.1$, $i$-band value of $L_*= 2.08\times 10^{10}h^{-2}\textup{L}_\odot$. To calculate the K-corrections, we make use of the `K-corrections calculator' Python algorithm\footnote{\url{http://kcor.sai.msu.ru/getthecode/}}, which is based on the procedure described in \citet{Chilingarian2010}.  

Once the values of $N_{\textup{1Mpc}}$ are determined, we calculate the cluster radius $R_{200}$, defined as the radius within which the cluster galaxy number density is $200\Omega_m^{-1}$ times the mean galaxy density of the present Universe. We calculate $R_{200}$ (in units of $h^{-1}$Mpc) using an empirical relation presented by \cite{Hansen2005}: 
\begin{equation}
 R_{200}= (0.142\pm0.004)N_{\textup{1Mpc}}^{0.6\pm0.01},
 \label{eq:R_200}
\end{equation}based on their analysis of the SDSS maxBCG clusters. $R_{200}$ is in turn used to obtain the final cluster richness $N_{200}$ which is calculated in the same way as $N_{\textup{1Mpc}}$, but now using $R_{200}$ as the aperture within which the cluster members are counted as opposed to the fixed aperture of 1 Mpc.

\subsection{Scaling $N_{200}$ by $n(z)$}
\label{sec:Cluster_richness_scaling}

Due to the magnitude limits of our data, our ability to detect fainter cluster members is reduced as a function of redshift, which could lead to an under-estimation of our cluster $N_{200}$ values at higher redshifts. In order to correct for the impact of the survey magnitude limits on our $N_{200}$ and subsequent cluster mass ($M_{\textup{200m}}$) estimation, we up-weight our $N_{200}$ values by a theoretical galaxy $n(z)$ curve which gives the relative number of galaxies detectable as a function of redshift, given our $i<21.1$ magnitude limit. This theoretical $n(z)$ is obtained based on a luminosity function which assumes a \cite{Schechter1976} function, with the i-band, `red' galaxy values of $\alpha=-0.46$ and the $z=0.1$ value of $M^*-5\log h= -20.63$ taken from Table 3 of \cite{Loveday2012}, $k$ and evolution corrected (with a star-formation timescale of $\tau=2.5$ Gyr), using the Stellar population synthesis model of \cite{Bruzual2003}.

\begin{figure*}
	\begin{subfigure}[hbt]{0.5\textwidth}
		\centering
        \caption{}
		\includegraphics[width=\textwidth]{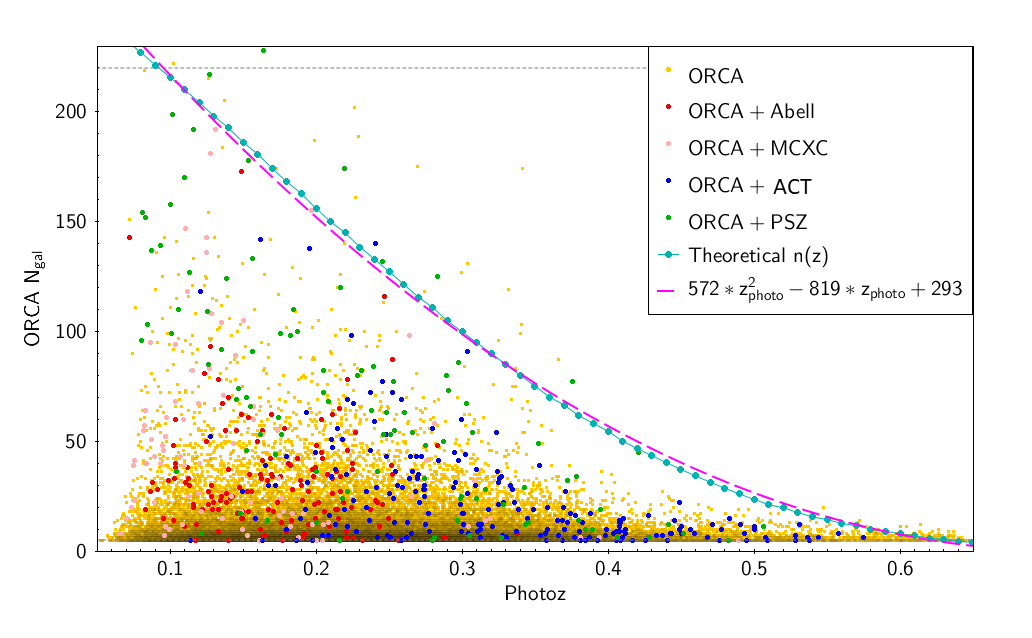}
		\label{fig:LF_scaling_1}
	\end{subfigure}\hspace{5mm} 
	\begin{subfigure}[hbt]{0.44\textwidth}
		\centering
		\caption{}
		\includegraphics[width=\textwidth]{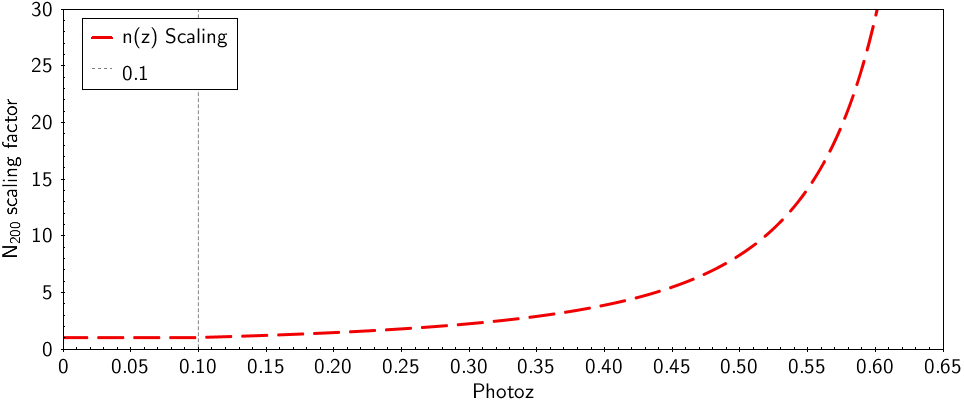}
		\label{fig:LF_correction_factor}
	\end{subfigure}
	\caption[Scaling cluster membership using the $n(z)$]{(a) {\sc orca} cluster membership $N_{gal}$ as a function of photometric redshift. We also highlight {\sc orca} clusters with counterparts in Planck SZ, MCXC, ACT DR5 and Abell cluster catalogues. The best fit curve (dashed pink line), fitted to the normalized theoretical $n(z)$ described in Section~\ref{sec:Cluster_richness_scaling} is used to determine the redshift dependent $N_{200}$ scaling factor shown in panel (b).}
	\label{fig:LF_scaling}
\end{figure*}

Figure~\ref{fig:LF_scaling_1} shows the cluster membership of our {\sc orca} clusters (highlighting {\sc orca} clusters matched to Planck, ACT DR5, MCXC and Abell cluster samples) as a function of redshift. Here, we are also plotting the theoretical $n(z)$ described above, normalized to match our maximum value of $N_{\textup{gal}}=220$ at $z=0.1$. Also shown is a fitted curve to the theoretical n(z) which is used to obtain the scaling factor,  $C(\bar{z})$, applied to our $N_{200}$ in the $z>0.1$ redshift range, where the number of our cluster members begin to drop as a function of redshift. This scaling factor takes the form of:
\begin{equation}
C(\bar{z})=220/(571.7\bar{z}^2-819\bar{z}+292.8),
\label{eq:nz_scaling}
\end{equation}i.e. the ratio of $N_{\textup{gal}}=220$ to our normalized theoretical $n(z)$ at each redshift and is shown in Figure~\ref{fig:LF_correction_factor} as a function of redshift.

\begin{figure*}
	\begin{subfigure}[hbt]{0.55\textwidth}
		\centering
		\caption{}
		\includegraphics[width=\textwidth]{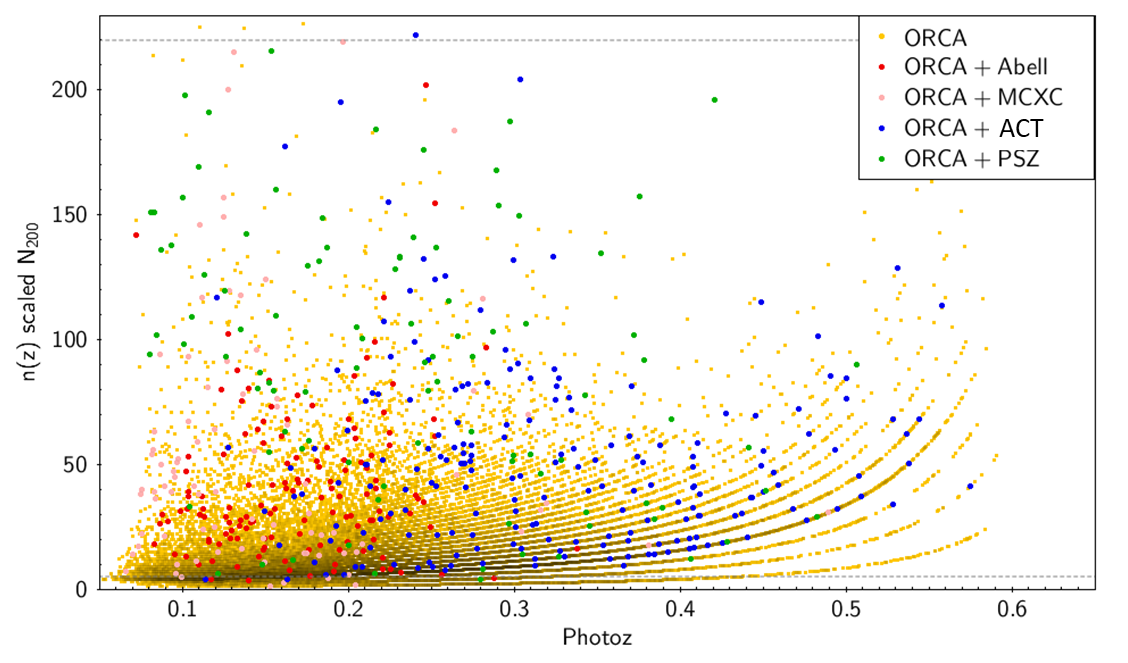}
            \label{fig:N_200_LF_scaled_all}
	\end{subfigure}
		\begin{subfigure}[hbt]{0.4\textwidth}
		\centering
		\caption{}
		\includegraphics[width=\textwidth]{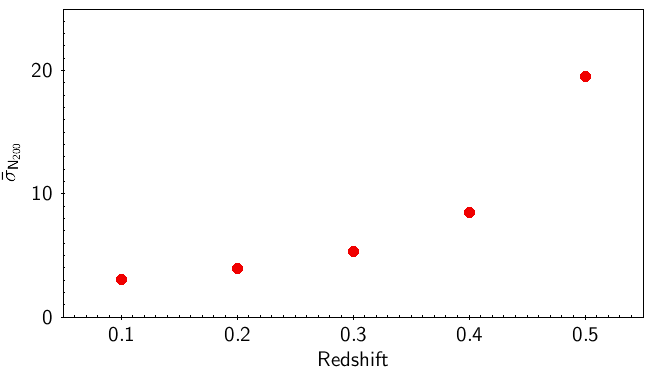}
		\label{fig:N200_err_vs_z}
	\end{subfigure}
	\caption[$n(z)$ scaled $N_{200}$ values as a function of photometric redshift]{(a) $n(z)$ scaled $N_{200}$ values as a function of photometric redshift. A comparison of this plot with Figure~\ref{fig:LF_scaling_1} shows that this scaling has compensated for our tendency to underestimate cluster richness values with increasing redshift. For clarity, we do not show the error bars on the scaled $N_{200}$ values in panel (a). However, in panel (b) we show the mean uncertainty on the scaled $N_{200}$ values as a function of cluster photometric redshift.}
	\label{fig:LF_scaling2}
\end{figure*}

In order to apply the $C(\bar{z})$ scaling to our $N_{200}$ values, we first scale our $N_{\textup{1Mpc}}$ values which are used to estimate the value of $R_{200}$ for each cluster, thus ensuring our $R_{200}$ values are not underestimated at higher redshifts. We then count the number of cluster galaxies within $R_{200}$ and scale this number by the $C(\bar{z})$ to obtain the $N_{200}$ values plotted in Figure~\ref{fig:N_200_LF_scaled_all}. In this manner, when estimating the richness of our clusters, we account for the increased likelihood of cluster galaxies remaining undetected with increasing redshift. The uncertainty on the scaled $N_{200}$ is then given by propagation of the $\sqrt{n}$ error on $N_{200}$ and the uncertainty on $C(\bar{z})$:

\begin{multline}
    \sigma_{N_{200}}=2200\times \\ \sqrt{\frac{4N_{200}^2(5717\bar{z}-4095)^2\sigma_{\bar{z}}^2+N_{200}(5717\bar{z}^2-8190\bar{z}+2928)^2}{(5717\bar{z}^2-8190\bar{z}+2928)^4}}.
    \label{eq:N_200_err}
\end{multline} Note that in the above equation the $N_{200}$ values are not scaled by $C(\bar{z})$. For the remainder of this work, however, unless otherwise specified, our use of $N_{200}$ refers to these $n(z)$ scaled $N_{200}$ values.

We note that our cluster mass estimates and the resulting mass functions as presented in Sections \ref{sec:ORCA_masses_results} and \ref{sec:arm_cluster_mfs} are not very sensitive to the chosen value of normalisation and exact parameters of equation~\ref{eq:nz_scaling}. This is because any systematic redshift-dependent offsets introduced during the $n(z)$ scaling is removed when we calibrate our cluster mass-richness scaling relation to cluster masses from external samples in the next Section.

\subsection{ATLAS cluster mass-richness scaling relation}
\label{sec:Cluster_mass}

Using our $N_{200}$ estimates from Section~\ref{sec:Cluster_richness}, along with the mean photometric redshifts of our clusters, $\bar{z}$, we provide cluster masses $M_{200\textup{m}}$ (in units of $10^{14} h^{-1}\textup{M}_\odot$) based on the following scaling relation:
\begin{equation}
 M_{200\textup{m}}=\left(\frac{N_{200}}{20}\right)^{1.1}\times3\bar{z}\ ^{0.9}+1.
\label{eq:M_200}
\end{equation}

\begin{figure*}
	\begin{subfigure}[t]{\columnwidth}
		\centering
		\caption{}
		\includegraphics[width=1.1\columnwidth]{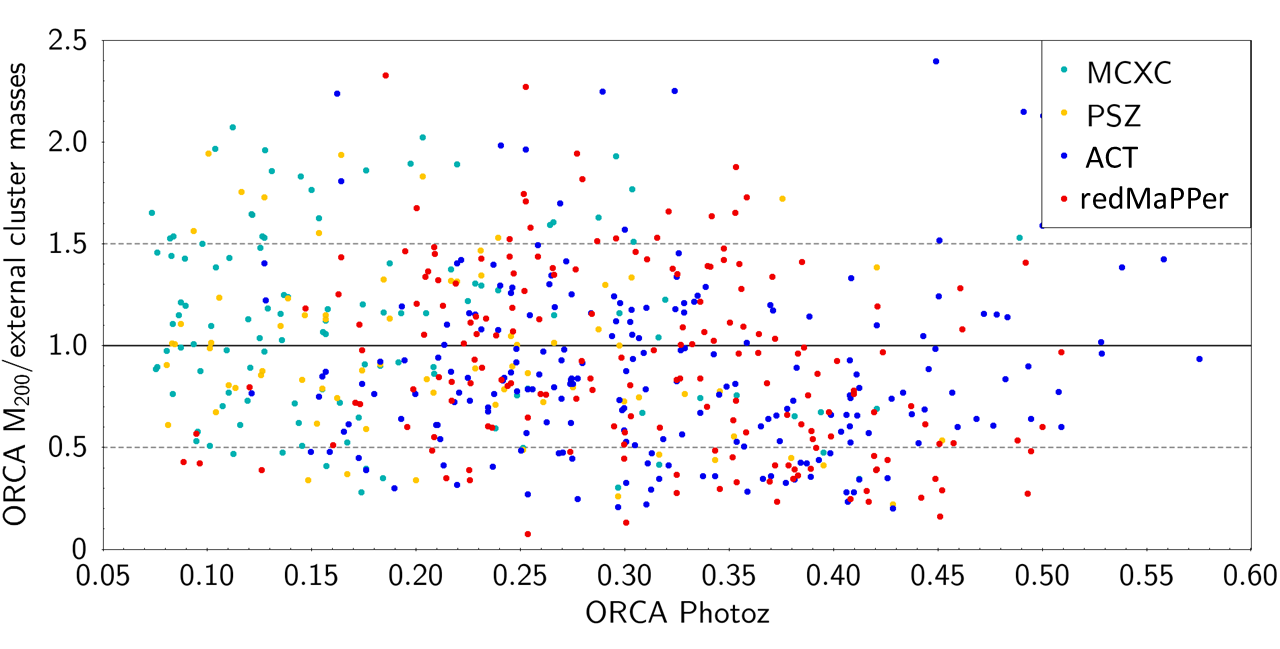}
		\label{fig:mass_ratios}
	\end{subfigure}\hfill%
	\begin{subfigure}[t]{\columnwidth}
		\centering
		\caption{}
		\includegraphics[width=0.9\columnwidth]{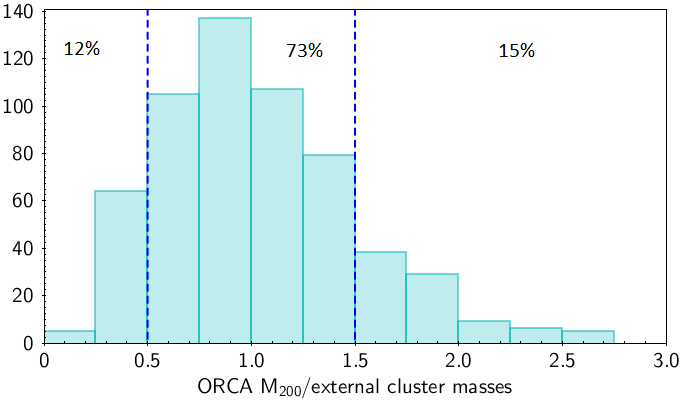}
        \label{fig:mass_ratio_hist}
	\end{subfigure}
	\caption[]{(a) The ratio of our {\sc orca} cluster masses to masses of their cluster counterparts from external samples. The dotted lines show regions where our masses are in agreement with external masses within $\pm50\%$. (b) Histogram showing the ratio of the {\sc orca} cluster masses to external cluster masses shown in (a). Here, we mark the percentage of clusters with mass ratios below, above and within the $\pm50\%$ region.}
	\label{fig:Cluster_mass_ratios}
\end{figure*}

The form and parameters of this relation are chosen based on comparison\footnote{The ATLAS mass-richness scaling relation is chosen to minimise the offset and scatter in Figure~\ref{fig:mass_ratios}.} of the resulting ATLAS cluster mass estimates to masses of $626$ clusters from external samples. These include $218$ ACT DR5 and $95$ Planck SZ clusters, $185$ SDSS redMaPPer clusters and $118$ MCXC X-ray clusters. The error on our cluster masses are then approximated using:

\begin{equation}
\sigma_{M_{200\textup{m}}}=\sqrt{0.015\sigma_{N_{200}}^2 N_{200}^{0.2} \bar{z}^{1.8}+\frac{0.01N_{200}^{2.2} \sigma_{\bar{z}}^2}{\bar{z}^{0.2}}},
\label{eq:M_200_err}
\end{equation}

\noindent where $\sigma_{\bar{z}}$ and $\sigma_{N_{200}}$ are the uncertainties on $\bar{z}$ and $N_{200}$ respectively.

\begin{figure*}
	\begin{subfigure}[hbt]{\columnwidth}
		\centering
		\caption{}
		\includegraphics[width=1\columnwidth]{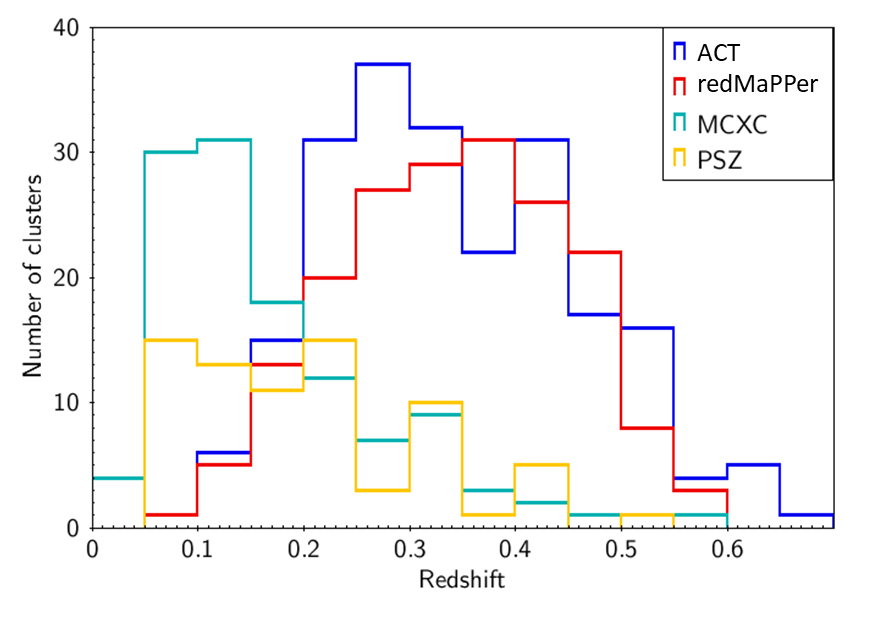}
		\label{fig:mass_calib_z_hist}
	\end{subfigure}\hspace{5mm} 
	\begin{subfigure}[hbt]{\columnwidth}
		\centering
		\caption{}
		\includegraphics[width=1.05\columnwidth]{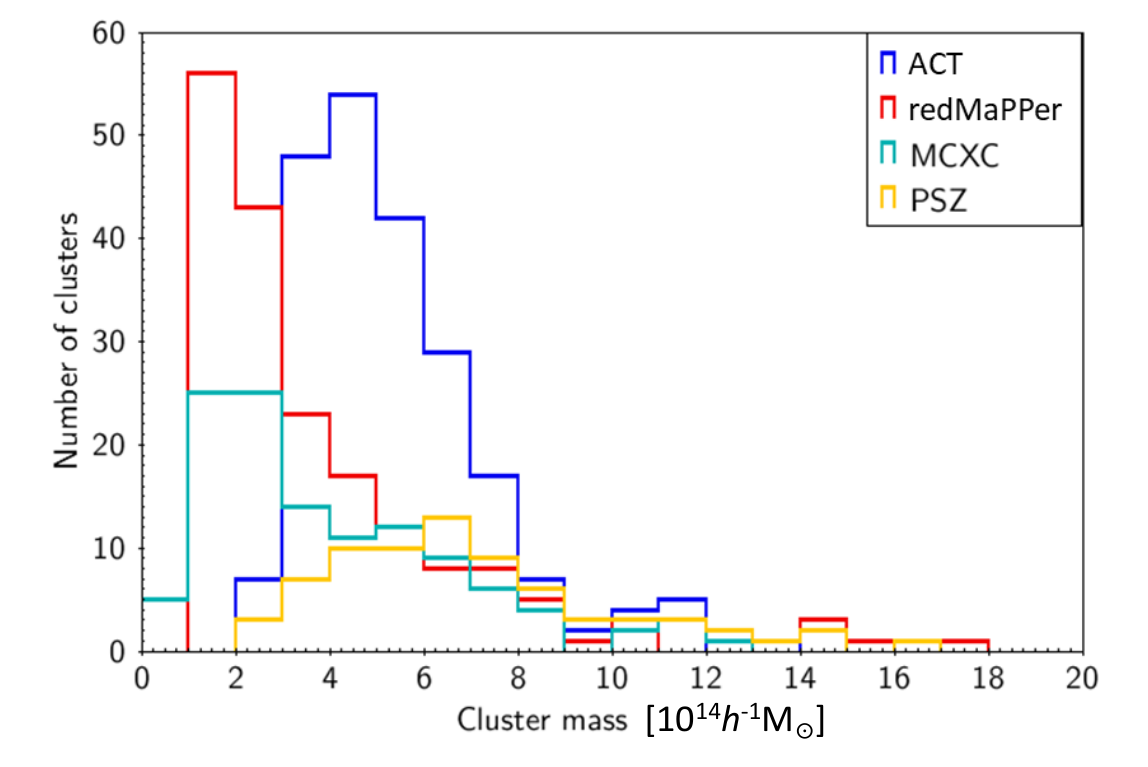}
		\label{fig:mass_calib_mass_hist}
	\end{subfigure}
	\caption[Redshift and mass distribution of the cluster samples used to calibrate ATLAS cluster masses]{Redshift (a) and mass (b) distributions of the clusters from the ACT DR5, redMaPPer, MCXC and Planck SZ cluster catalogues, used to calibrate the ATLAS cluster masses.}
	\label{fig:mass_z_calib_hists}
\end{figure*} 

Figure~\ref{fig:mass_z_calib_hists} shows the redshift and mass distribution of the clusters from these four samples. The combination of these samples provide a reasonable number of counterparts to ATLAS clusters across the redshift range of $0.05<z<0.55$ and the mass range of $10^{14}-10^{15}$ M\textsubscript{\(\odot\)}. To select the cluster counterparts, we use the {\sc orca} cluster radius ($R_{\rm ORCA}$) as our matching radius for the ACT DR5, redMaPPer, MCXC and Abell cluster catalogues, while using the Planck position uncertainty on the SZ cluster centre as the matching radius. Here, we calibrate the parameters in our cluster mass-richness scaling relation (Equation~\ref{eq:M_200}) using all available cluster masses from these four samples simultaneously, to avoid biasing our masses towards the range of masses covered by the individual samples. 

We obtain the masses of the redMaPPer clusters using the weak lensing calibrated mass-richness scaling relation of \cite{Simet2016}, with $M_{\textup{200m}}=10^{14.344}(\lambda/40)^{1.33}$. Here, $M_{\textup{200m}}$ is the $M_{200}$ cluster mass with respect to the mean density of the universe (in units of $10^{14}h^{-1}$\(\textup{M}_\odot\)) and $\lambda$ is the richness of redMaPPer clusters as defined by \cite{Rykoff2014}. Similarly, we use the $M_{\textup{200m}}$ cluster masses from the ACT DR5 catalogue. As the MCXC cluster masses are defined as $M_{500}$, we multiply these by a factor of 1.5 in order to remove the mean offset found between the MCXC $M_{500}$ masses and the $M_{\textup{200m}}$ masses of their redMaPPer and ACT DR5 counterparts. We correct for the mean offset between the Planck SZ mass proxy values, $M_{\textup{SZ}}$, and the $M_{\textup{200m}}$ masses of their redMaPPer and ACT DR5 counterparts, by multiplying the Planck masses by a factor of 1.3. Once the Planck and MCXC masses are scaled to roughly correspond to $M_{\textup{200m}}$ masses, we use the masses from the four samples to obtain the ATLAS mass-richness scaling relation given by Equation~\ref{eq:M_200}. In the case of the ACT DR5 and Planck SZ samples, we limit the sources to those with a SNR>5.

The ratio of our {\sc orca} cluster masses to masses based on their counterparts from these four samples are shown in Figure~\ref{fig:mass_ratios}. Our cluster masses are calibrated on the basis of minimizing the mean offset and scatter in this plot as well as the mass ratio histogram shown in Figure~\ref{fig:mass_ratio_hist}. The level of scatter found in the ratio of our richness-based cluster mass estimates to masses from external X-ray and SZ samples is comparable to that found when we compare the redMaPPer richness-based cluster masses to masses from these external samples.

The photometric redshift term in our ATLAS mass-richness scaling relation (Equation~\ref{eq:M_200}) is added to remove a redshift dependent offset found between our $M_{\textup{200m}}$ masses and the masses from the four external catalogues. This redshift dependence may be partially due to the presence of a slight systematic offset towards lower redshifts in our photometric redshift estimates in the $z>0.35$ redshift range, which is likely caused by the relatively lower number of galaxies in our spectroscopic training set in this redshift range (see Figures~\ref{fig:spec_training_hist.png},~\ref{fig:wm_ext_comp_v1.pdf} and the discussion in Section~\ref{sec:Photoz_results}). The presence of such systematic offset towards lower redshift will, in turn, result in a lower $n(z)$ scaling factor being applied to the high redshift cluster richness values, biasing our estimates of $N_{\textup{200}}$ and $M_{\textup{200m}}$ low.

In addition, although no evidence has been found to suggest cluster mass-to-light ratios ($M/L$) increase as a function of redshift (e.g. \citealt{Lin2006}; \citealt{Muzzin2007}; \citealt{Soucail2015}), the increase of cluster $M/L$ as a function of cluster mass has been well established. Over a wide range of masses, various studies have found a ratio of $M/L\propto M^{\alpha}$, with $\alpha$ ranging from $0.2-0.4$ (\citealt{Girardi2002}; \citealt{Lin2003}; \citealt{Rines2004}; \citealt{Popesso2007}). Given that with increasing redshift (and in the $z>0.3$ range in particular) due to the magnitude limits of our sample we are increasingly likely to miss smaller, lower mass clusters, with the same being true (to varying degrees) for the external cluster samples used in our mass calibrations\footnote{This is due to the flux limit of the redMaPPer and MCXC samples, and for Planck and ACT DR5, due to the SZ lower-limit of cluster mass observability.}, the increased fraction of higher mass clusters at higher redshifts could mean that our $N-M$ scaling relation which provided a good estimate of total cluster masses at lower redshifts, is under-estimating the mass of our high redshift population which tend to have higher $M/L$ ratios due to their higher mass.

\section{Results and Discussion}
\label{sec:Results}

\subsection{The VST ATLAS cluster catalogue}
\label{sec:Cat_cols}

Table~\ref{tab:cat_centre_cols} provides a description of the ATLAS catalogue columns, detailing the various properties of clusters including redshift, richness, cluster radius and cluster mass. For each catalogue column, we provide the column title given in the catalogue, the corresponding symbol as used in the text of this work, followed by the units and a description of the values in each column. Table~\ref{tab:cat_members_cols} provides a description of the columns in the ATLAS cluster members catalogues providing unique identifiers, coordinates and photometric redshifts of each cluster galaxy.

\begin{table*}
    \caption{A description of the columns of the ATLAS cluster catalogue. For each catalogue column, the symbols column of this Table shows the corresponding symbol used in the text and figures of this paper.}
    \label{tab:cat_centre_cols}
    \begin{tabular}{l|c|p{13cm}}
    \hline
    Column & Symbol [units] & Description \\
        \hline
         ClusterID & --- & A unique cluster identification number assigned to each cluster by the {\sc orca} cluster detection algorithm.\\[2mm]
         RA & [Degrees J2000] & Right Ascension of the cluster centre (mean RA of cluster members).\\[2mm]
         DEC & [Degrees J2000] & Declination of the cluster centre (mean Dec of cluster members).\\[2mm]
         Photoz & $\bar{z}$ & Error-weighted mean photometric redshift of the cluster.\\[2mm]
         Photoz\_err & $\sigma_{\bar{z}}$ & Standard Error on the mean cluster photometric redshift.\\[2mm]
         R\_ORCA & $R_{\rm ORCA}$ [arcmin] & The projected radius of the {\sc orca} cluster on the sky, defined as the angular separation between the cluster centre and the furthest cluster galaxy.\\[2mm]
         ORCA\_Ngal & $N_{gal}$ & The number of cluster galaxies detected by {\sc orca}.\\[2mm]
         N\_1Mpc & $N_{1Mpc}$ & The number of cluster galaxies detected by {\sc orca} within a radius of $1 h^{-1}$Mpc from the centre of the cluster. This number is scaled by the theoretical $n(z)$ following Equation~\ref{eq:nz_scaling}.\\[2mm]
         R\_200 & $R_{200}$ [arcmin] & The radius from the cluster centre within which the density is $200\times$ the mean density of the Universe (given by Equation~\ref{eq:R_200}).\\[2mm]
         N\_200 & $N_{200}$ & The number of cluster galaxies within a radius of $R_{200}$ from the centre of the cluster. This number is scaled by the theoretical $n(z)$ following Equation~\ref{eq:nz_scaling}.\\[2mm]
         N\_200\_err & $\sigma_{N_{200}}$ & The error on $N_{200}$ given by Equation~\ref{eq:N_200_err}.\\[2mm]
         M\_200m & $M_{200\textup{m}}$ [$10^{14} h^{-1}\textup{M}_\odot$] & The cluster mass enclosed within a radius of $R_{200}$ from the centre of the cluster (given by Equation~\ref{eq:M_200}). The cluster masses in our catalogue are measured with respect to the mean density of the Universe (often represented using the symbol $M_{200m}$ in literature). \\[2mm]
         M\_200m\_err & $\sigma_{M_{200\textup{m}}}$ [$10^{14} h^{-1}\textup{M}_\odot$] & The uncertainty on the $M_{200\textup{m}}$ cluster mass given by Equation~\ref{eq:M_200_err}.\\[2mm]
         \hline
         \end{tabular}
\end{table*}

\begin{table*}
    \centering
    \caption[Description of the columns in the ATLAS cluster members catalogue]{A description of the columns of the ATLAS cluster members catalogue. For each catalogue column, the symbols column of this Table shows the corresponding symbol used in the text and figures of this paper.}
    \begin{tabular}{l|c|p{13cm}}
        \hline
        Column & Symbol [units] & Description \\
         \hline
         ClusterID & --- & A unique cluster identification number assigned to each cluster by the {\sc orca} cluster detection algorithm.\\[2mm]
         ObjID & --- & A unique object identification number assigned to each cluster galaxy by the {\sc orca} cluster detection algorithm.\\[2mm]
         RA & [Degrees J2000] & Right Ascension of the cluster member.\\[2mm]
         DEC & [Degrees J2000] & Declination of the cluster member.\\[2mm]
         g\_mag & $g_{\textup{A5}}$ & VST ATLAS $g$-band Aperture 5 (aperture radius of $2''$) magnitude in the AB system.\\[2mm]
         r\_mag & $r_{\textup{A5}}$ & VST ATLAS $r$-band Aperture 5 (aperture radius of $2''$) magnitude in the AB system.\\[2mm]
         i\_mag & $i_{\textup{A5}}$ & VST ATLAS $i$-band Aperture 5 (aperture radius of $2''$) magnitude in the AB system.\\[2mm]
         z\_mag & $z_{\textup{A5}}$ & VST ATLAS $z$-band Aperture 5 (aperture radius of $2''$) magnitude in the AB system.\\[2mm]
         W1\_mag & W1 & unWISE W1 magnitude in the AB system.\\[2mm]
         W2\_mag & W2 & unWISE W2 magnitude in the AB system.\\[2mm]
         Photoz & $z$ & Cluster galaxy photometric redshift as determined by {\sc ANNz2} machine learning algorithm (see Section~\ref{sec:ANNz2}).\\[2mm]
         Photoz\_err & $\sigma(z)$ & Error on cluster galaxy photometric redshift as determined by {\sc ANNz2}.\\[2mm]
         \hline
    \end{tabular}
    \label{tab:cat_members_cols}
\end{table*}

\subsection{Photometric redshifts}
\label{sec:Photoz_results}

\begin{figure*}
	\begin{subfigure}[T]{\columnwidth}
		\centering
		\caption{}
		\includegraphics[width=0.8\columnwidth]{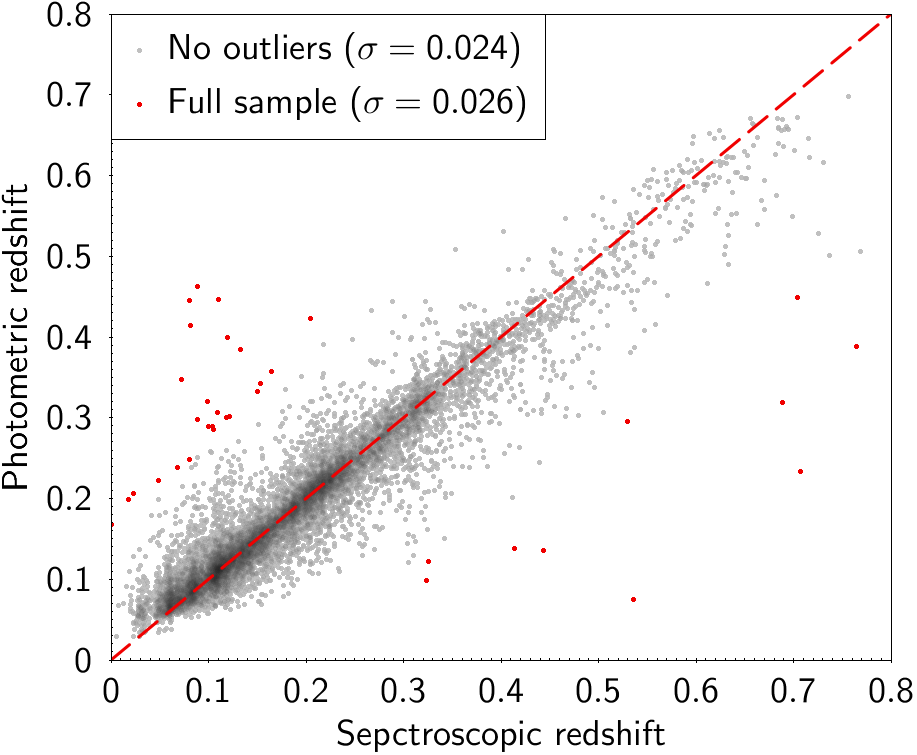}
		\label{fig:Photoz_rms.pdf}
	\end{subfigure}	
	\begin{subfigure}[T]{\columnwidth}
		\centering
		\caption{}
		\includegraphics[width=0.8\columnwidth]{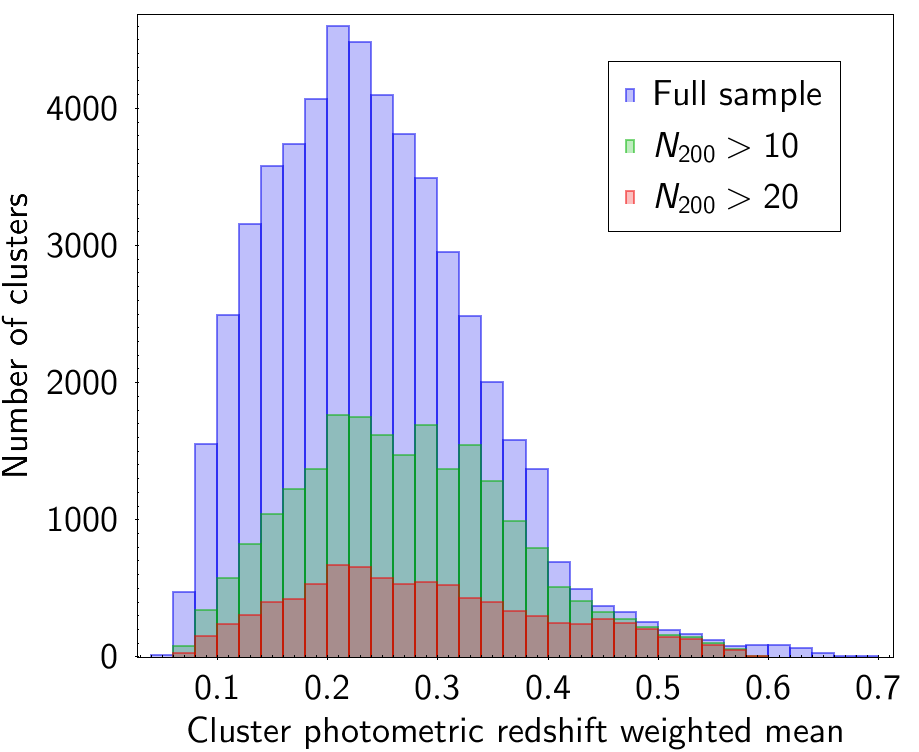}
		\label{fig:Mean_clusters_photozs.pdf}
	\end{subfigure}
	\begin{subfigure}[T]{\columnwidth}
		\centering
		\caption{}
		\includegraphics[width=0.95\columnwidth]{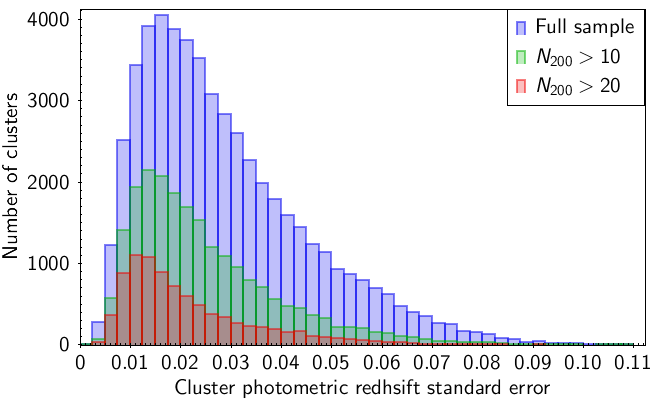}
		\label{fig:Mean_clusters_photoz_errs.pdf}
	\end{subfigure}
	\begin{subfigure}[T]{\columnwidth}
		\centering
		\caption{}
		\includegraphics[width=0.8\columnwidth]{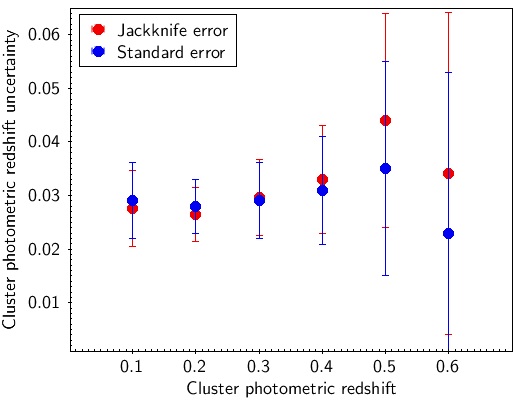}
		\label{fig:wm_err_vs_wm.png}
	\end{subfigure}
	\caption[VST ATLAS cluster photometric redshifts]{(a) Photometric vs. spectroscopic redshift of the {\sc orca} cluster members. The photometric redshifts are calculated following the procedure described in Section~\ref{sec:ANNz2}. Outliers (red) are defined as $(|z_{\rm photo}-z_{\rm spec}|/(1+z_{\rm spec}))>0.15$. (b) The distribution of the (error-weighted) mean photometric redshift of the VST ATLAS clusters, with the standard error on the cluster photometric redshift shown in panel (c).  (d) A comparison of the Jackknife and standard error estimates of the cluster photometric redshift uncertainties shows a good agreement between the two estimates. Here the error bars represent the error in the error given by $1/\sqrt{2N-2}$ where $N$ is the number of clusters in each redshift bin.}
	\label{fig:Cluster_photozs}
\end{figure*}

\begin{figure}
    \begin{subfigure}[t]{\columnwidth}
    \centering
    \includegraphics[width=\columnwidth]{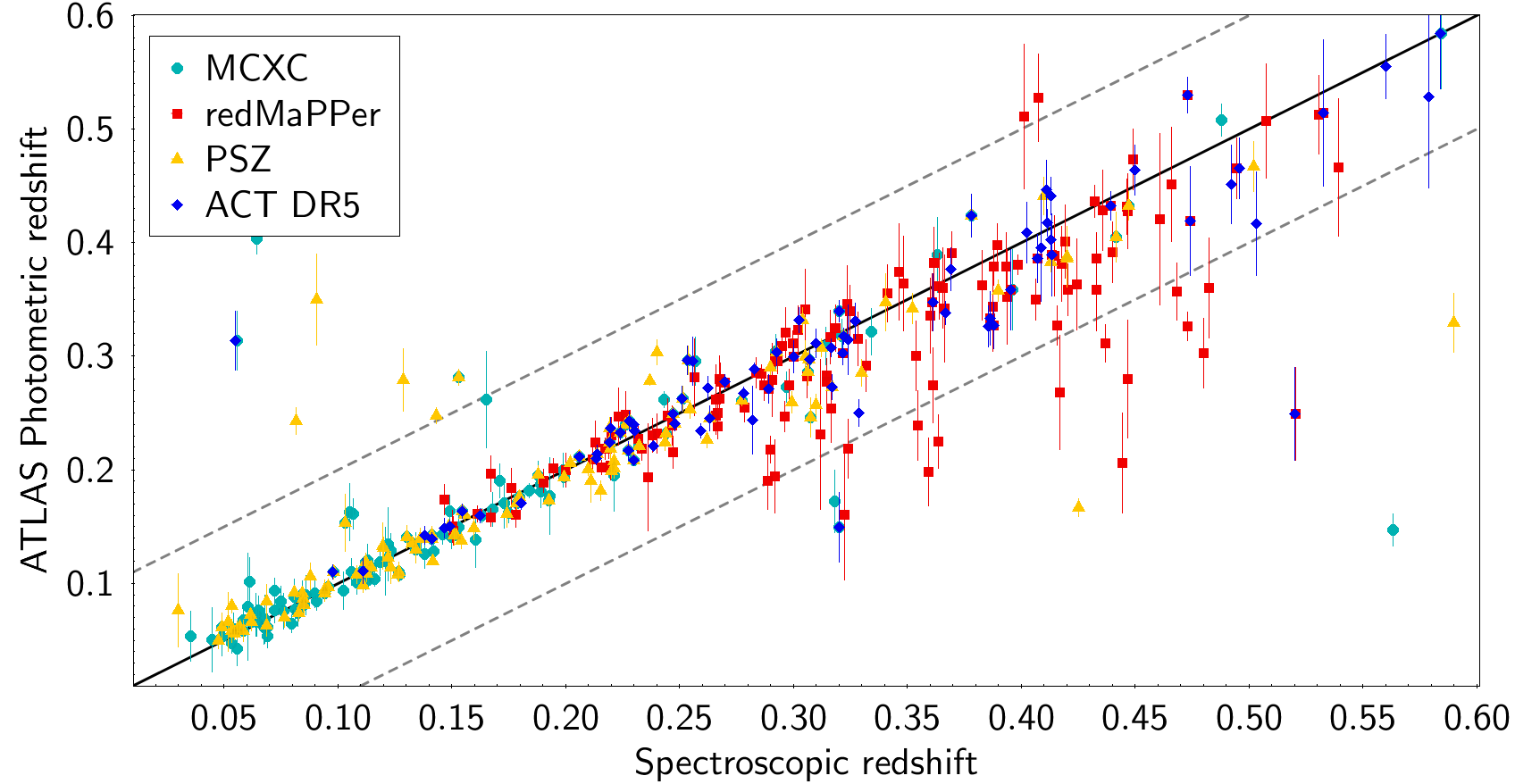}
    \end{subfigure}
    \caption[Comparison of ATLAS cluster photometric redshifts to spectroscopic redshifts of external X-ray, SZ and optical cluster detections]{A comparison of our ATLAS cluster photometric redshifts to spectroscopic redshifts from the MCXC, SDSS redMaPPer, Planck and ACT DR5 SZ cluster samples. The dotted lines mark the $\pm0.1$ ($\sim3\sigma$) error region.}
    \label{fig:wm_ext_comp_v1.pdf}
\end{figure}

Following the procedure described in Section~\ref{sec:ANNz2}, we use ANNz2 to obtain photo-z estimates for our cluster galaxies with an RMS scatter of $\sim0.024$ (see Figure~\ref{fig:Photoz_rms.pdf}), upon removing the outliers given by $((z_{\rm photo}-z_{\rm spec})/(1+z_{\rm spec}))>0.15$. Inclusion of outliers results in an RMS of $\sim0.026$. The error-weighted mean photometric redshift of the full ATLAS sample containing $\sim40,000$ {\sc orca} detections with $N_{200}\geq5$, as well as $\sim22,000$ clusters with $N_{200}>10$ and $9,000$ clusters with $N_{200}>20$ are shown in Figure~\ref{fig:Mean_clusters_photozs.pdf}. Here, the peaks of the distributions lie at $z\sim0.25$ and clusters are detected up to $z\sim0.7$, (or $z\sim0.6$ for clusters with $N_{200}>20$).

Figure~\ref{fig:Mean_clusters_photoz_errs.pdf} shows the distribution of the Standard error on the mean cluster redshift peaking at $\sim0.02$ for the full sample, and $\sim0.01$ for clusters with $N_{200}>20$. Finally, for the full sample, we show the uncertainty on the mean cluster photometric redshifts as a function of redshift in Figure~\ref{fig:wm_err_vs_wm.png}, finding a good agreement between the Jackknife and standard error estimates of the uncertainty.

In Figure~\ref{fig:wm_ext_comp_v1.pdf} we compare the photometric redshift of our ATLAS clusters with spectroscopic redshifts from their cluster counterparts from MCXC, SDSS redMaPPer, Planck and ACT DR5 samples. Here the $>3\sigma$ outliers (based on mean cluster photo-z error of $\sim0.03$) are marked by the dotted lines and constitute $\sim6\%$ of the sample. While we find a general agreement between the ATLAS photometric redshifts and the spectroscopic redshifts from external samples, we also note hints of systematics at $z>0.3$ where our photometric redshifts appear more likely to be under-estimated. 

\subsection{Mass and redshift completeness}
\label{sec:Mass_redshift_completeness}
 
\begin{figure}
	\begin{subfigure}[T]{\columnwidth}
		\centering
		\caption{Relative redshift completeness}
		\includegraphics[width=0.8\columnwidth]{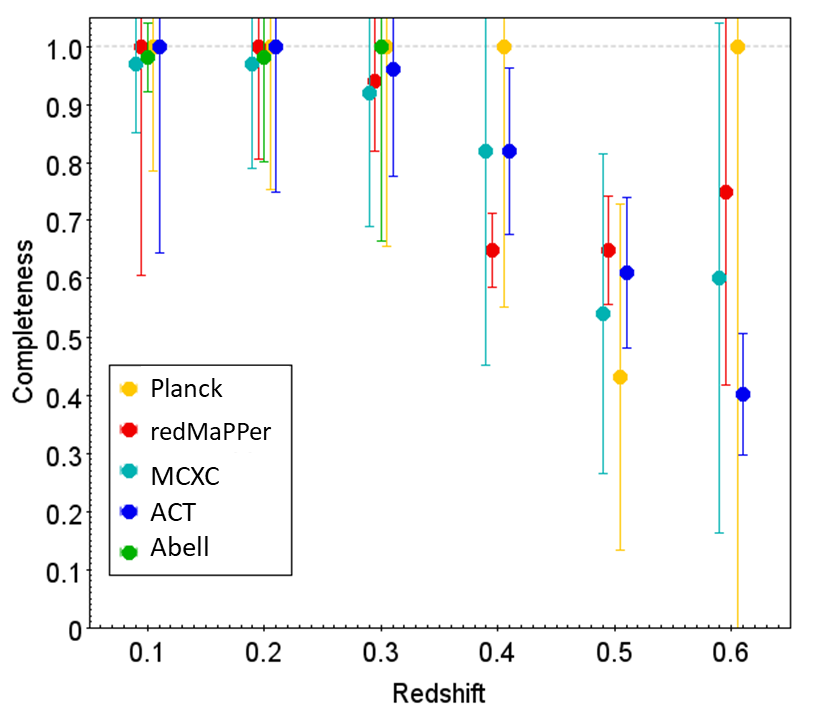}
		\label{fig:redshift_completeness_all}
	\end{subfigure}
	\begin{subfigure}[T]{\columnwidth}
		\centering
		\caption{Mean redshift completeness}
		\includegraphics[width=0.8\columnwidth]{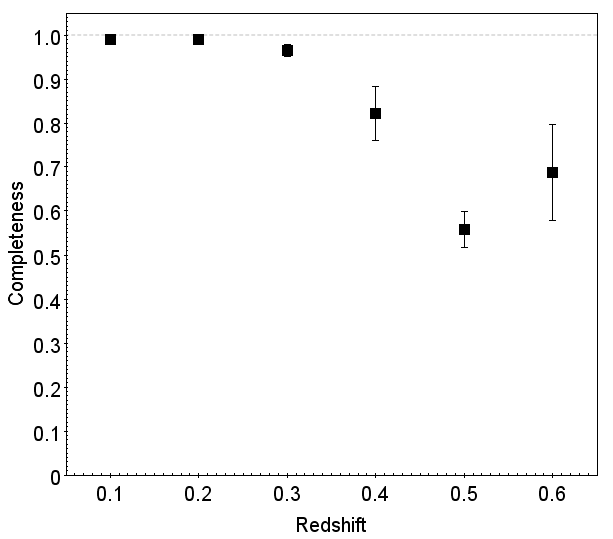}
		\label{fig:Mean_redshift_completeness}
	\end{subfigure}
	\caption[Redshift completeness of the ATLAS cluster sample]{(a) The redshift completeness of the ATLAS cluster sample, based on the fraction of the recovered SDSS redMaPPer, ACT DR5, Planck, MCXC and Abell clusters overlapping the ATLAS coverage area (see text for selection and matching criteria). The error bars are given by the propagation of $\sqrt{n}$ error estimates and for clarity, the data points corresponding to different datasets have been slightly shifted along the x-axis. (b) The mean redshift completeness of the ATLAS clusters sample based on the comparison to external clusters in panel (a). Here, the error bars are given by the standard error on the mean.}
	\label{fig:ORCA_redshift_completeness}
\end{figure}

Figure~\ref{fig:redshift_completeness_all} shows the fraction of clusters from the SDSS redMaPPer, ACT DR5, Planck, MCXC and Abell catalogues overlapping the ATLAS coverage area, that are detected in the ATLAS cluster catalogue. This provides a measure of the completeness of the ATLAS cluster sample as a function of redshift. In the case of the ACT DR5 and Planck samples, we limit the match to clusters with SZ detections with $SNR>5$. In all cases, the cluster samples are limited to clusters with masses greater than $1\times10^{14}h^{-1}$\(\textup{M}_\odot\), roughly corresponding to the lower mass limit of the ATLAS clusters. Figure~\ref{fig:Mean_redshift_completeness} shows the mean completeness of the ATLAS cluster samples as a function of redshift, where the sample is $>95\%$ complete in the range $z<0.3$ and $>80\%$ complete up to $z=0.4$. We note that the sharp fall in our completeness comparison to Planck at $z=0.5$ is due to small number statistics of the Planck sample at this redshift range where we detect 3/7 Planck SZ clusters overlapping the ATLAS coverage area.  

\begin{figure}
	\begin{subfigure}[T]{\columnwidth}
		\centering
		\caption{Relative cluster mass completeness}
		\includegraphics[width=0.82\columnwidth]{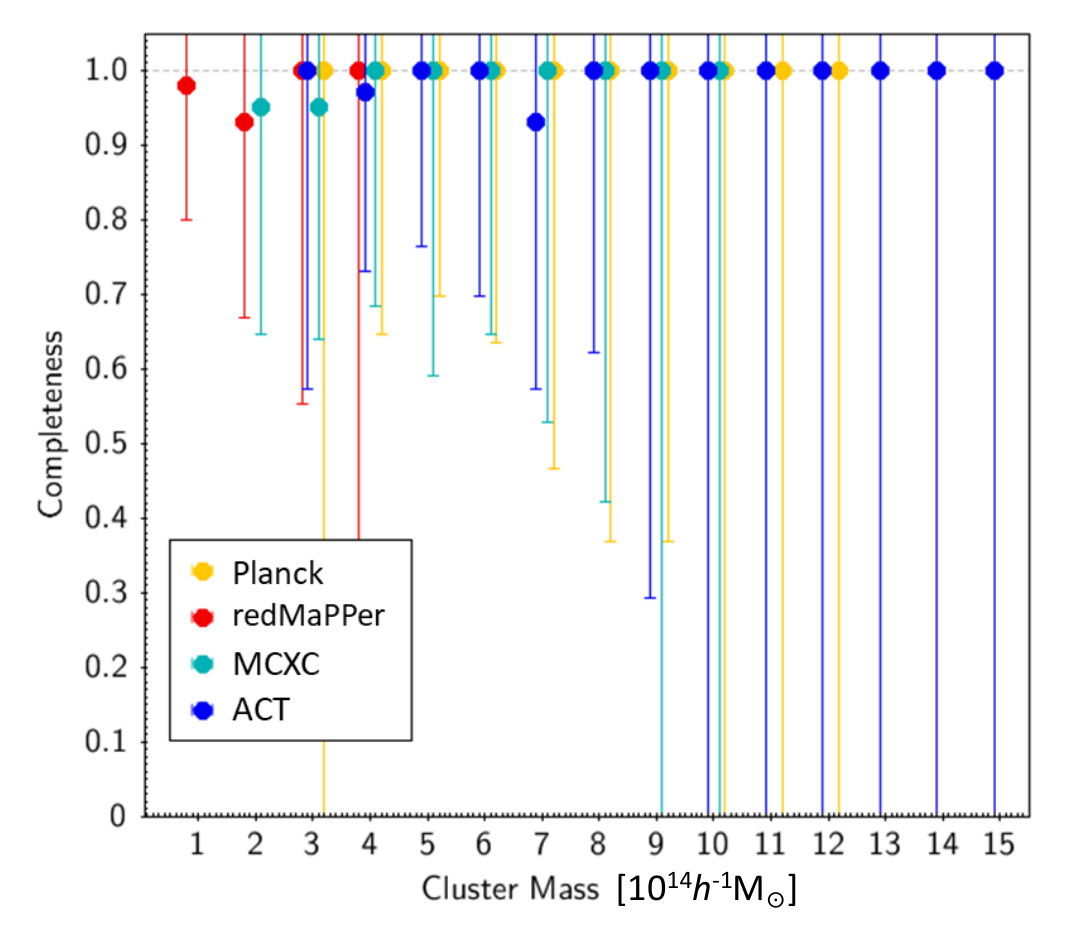}
		\label{fig:mass_completeness_all}
	\end{subfigure}
	\begin{subfigure}[T]{\columnwidth}
		\centering
		\caption{Mean cluster mass completeness}
		\includegraphics[width=0.85\columnwidth]{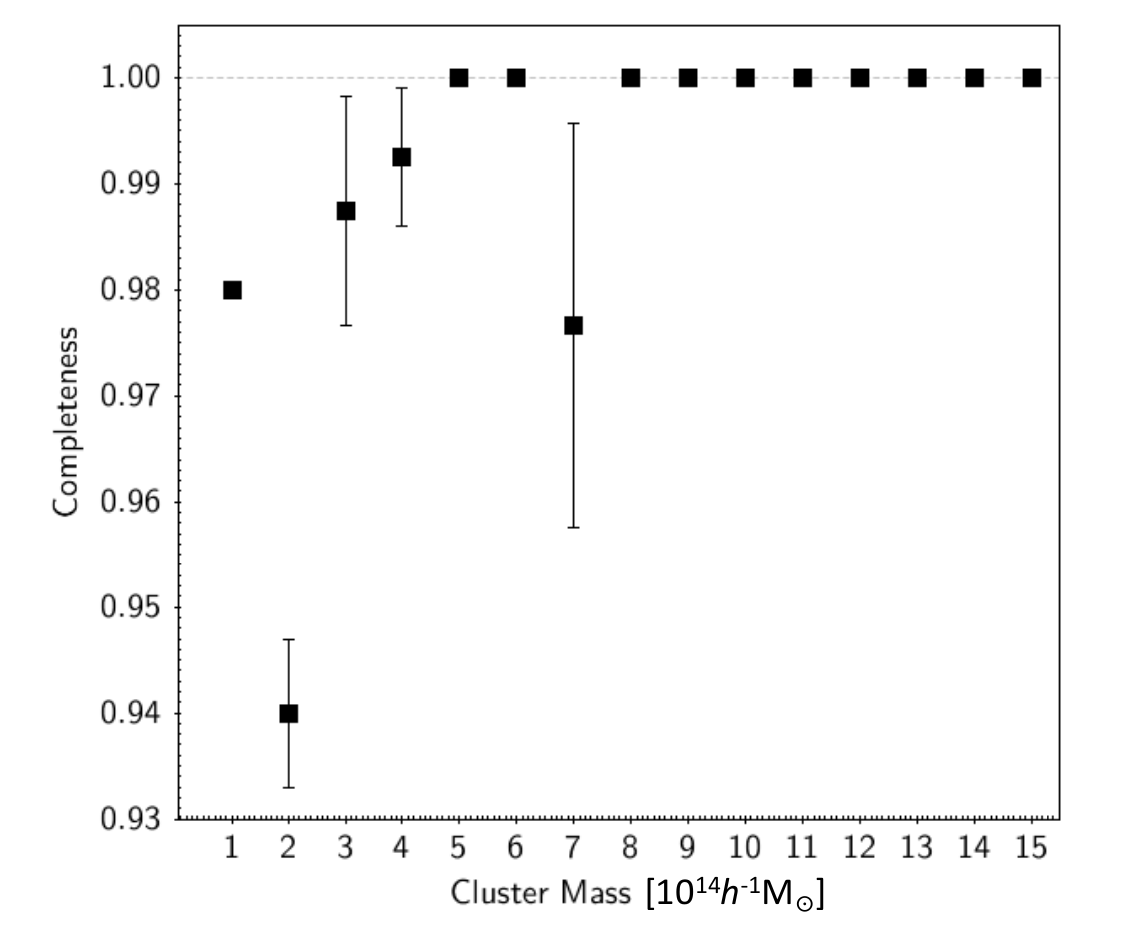}
		\label{fig:Mean_mass_completeness}
	\end{subfigure}
	\caption[Mass completeness of the ATLAS cluster samples]{(a) The fraction of SDSS redMaPPer, ACT DR5, Planck and MCXC clusters overlapping the ATLAS coverage area which are detected in the ATLAS cluster catalogue. This provides an estimate of the completeness of the ATLAS cluster sample as a function of mass. The error bars shown in this panel are given by the propagation of $\sqrt{n}$ error estimates and for clarity, the data points corresponding to different datasets have been slightly shifted along the x-axis. (b) The mean completeness of the ATLAS sample as a function of cluster mass, based on comparison to the samples shown in panel (a). Here, the error bars are given by the standard error on the mean.}
	\label{fig:ORCA_mass_completeness}
\end{figure}

Figure~\ref{fig:mass_completeness_all} shows the mass completeness of the ATLAS cluster catalogue by assessing the fraction of SDSS redMaPPer, ACT DR5, MCXC and Planck clusters detected in the ATLAS sample as a function of cluster mass. Here, all samples are limited to the redshift range $0.1<z<0.3$ in order to ensure the mass completeness is not impacted by our reduced completeness at higher redshifts. Figure~\ref{fig:Mean_mass_completeness} shows the mean cluster mass completeness of the ATLAS sample, with the sample being $>95\%$ complete across the full $1\times10^{14}-1.5\times10^{15}h^{-1}$\(\textup{M}_\odot\) mass range, with a near full recovery rate of external clusters for masses $>5\times10^{14}h^{-1}$\(\textup{M}_\odot\).

\subsection{Comparison to redMaPPer}
\label{sec:RM_comparison}

\begin{figure}
	\begin{subfigure}[t]{\columnwidth}
		\centering
		\includegraphics[width=\columnwidth]{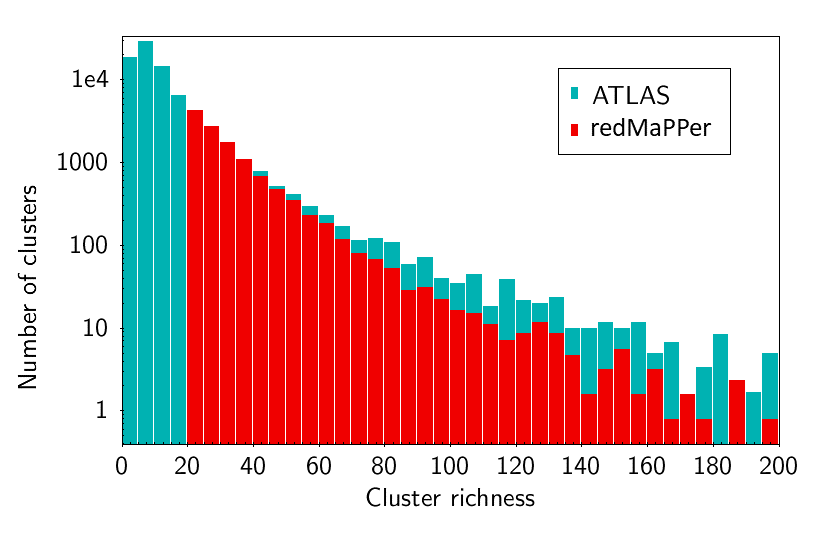}
	\end{subfigure}
	\caption[Cluster group size histograms]{ATLAS cluster richness $N_{200}$ distribution in comparison to SDSS redMaPPer cluster richness $\lambda$. Here both samples are limited to the redshift range $0.1<z<0.4$, and the redMaPPer histogram is scaled down by a factor of 2.2 to account for the differences between survey areas.}
	\label{fig:Cluster_hists}
\end{figure}

Figure~\ref{fig:Cluster_hists} shows the number of ATLAS clusters as a function of cluster richness ($N_{200}$). A similar histogram showing the richness ($\lambda$) of the SDSS redMaPPer sample is also added for comparison. Although redMaPPer clusters with ($\lambda<20$) are not available with the public release of the catalogue, it can be seen that both samples follow similar cluster richness distributions with thousands of clusters in bins of richness smaller than $\sim40$, hundreds in richness bins between $\sim40$ and 80, and tens of clusters in bins of richness ranging from $\sim80-140$. The ATLAS sample however contains a slightly larger number of richer clusters ($\sim10\%$), which is likely due to differences in cluster detection algorithms and the definitions of cluster richness between the two samples. 

For the remainder of this section, we limit the ATLAS and redMaPPer samples to their mutual survey overlap area of $\sim200$ deg$^{2}$, which constitutes $\sim4\%$ and $\sim2\%$ of the total ATLAS and SDSS survey areas respectively. As the overlap area is situated at the edge of both surveys, we remove any cluster that lies within $5'$ of the survey boundaries. We highlight that given the limited overlap area, one should keep in mind that the comparisons in this section may not be representative of the complete cluster samples. In addition, unless otherwise specified, we limit the cluster samples to the $M_{\textup{200m}}$>$3\times10^{14}h^{-1}$\(\textup{M}_\odot\) mass range in the comparisons performed in this section.

\begin{figure*}
		\includegraphics[width=\textwidth]{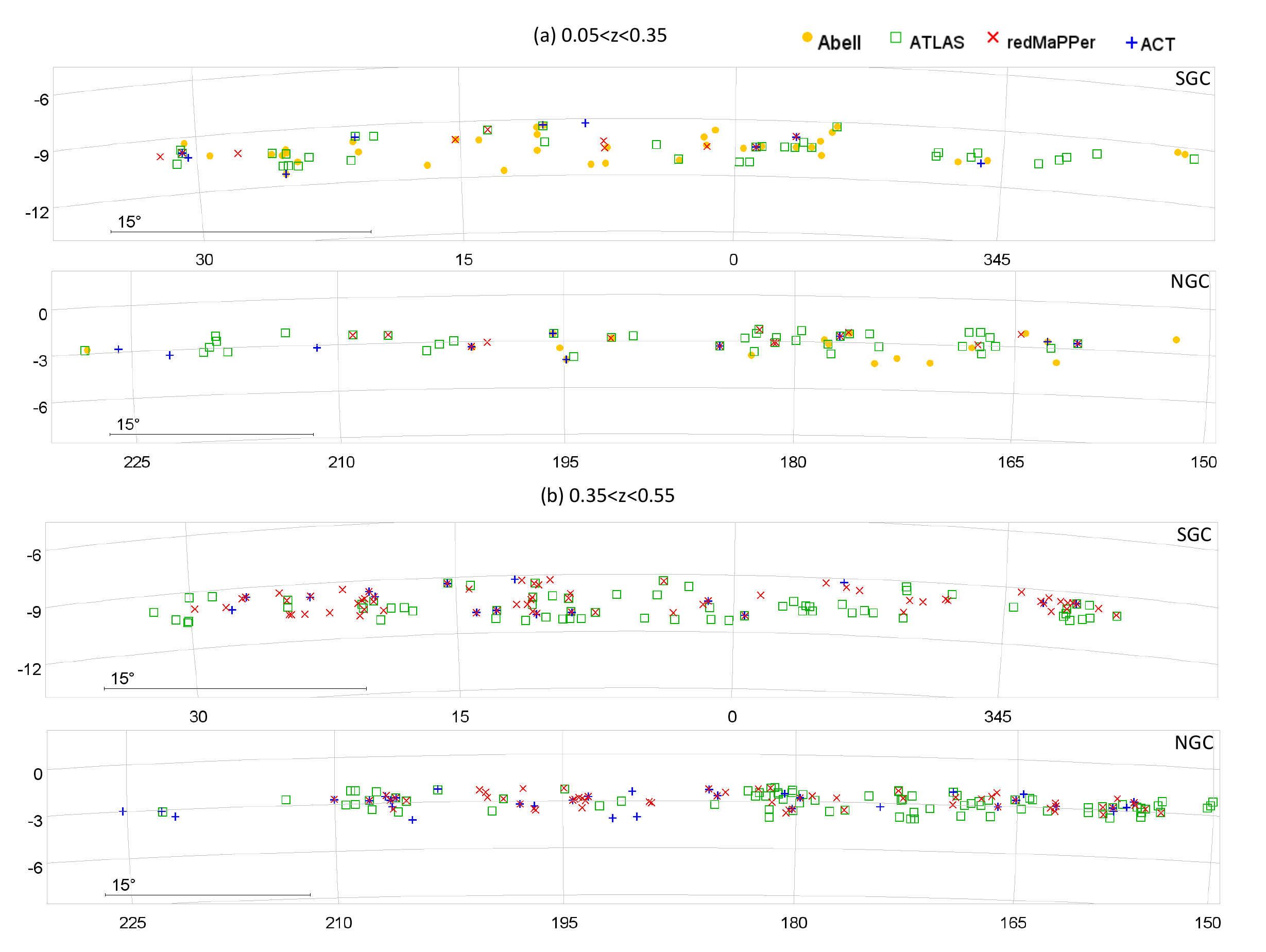}
	\caption[Comparison of ATLAS and redMaPPer catalogues in their overlap areas]{(a) Comparison of ATLAS, redMaPPer and Abell clusters limited to the redshift range $0.05<z<0.35$ in the overlapping areas of ATLAS and SDSS. (b) Same as (a) but now ATLAS and redMaPPer are compared to ACT DR5 clusters with all samples limited to the redshift range $0.35<z<0.55$. In all cases, ATLAS, redMaPPer and ACT clusters are limited to the $M_{\textup{200m}}$>$3\times10^{14}h^{-1}$\(\textup{M}_\odot\) mass range. We also note that the number density of Abell clusters in the ATLAS-SDSS overlap area in the NGC is $\sim50\%$ lower compared to the SGC.}
	\label{fig:arm_extcat_map}   
\end{figure*}

In Figure~\ref{fig:arm_extcat_map}, we show a comparison of the ATLAS, redMaPPer, Abell and ACT DR5 cluster catalogues in the ATLAS/SDSS survey overlap areas. For reasons that we will explore in more detail later, in the redshift range $0.05<z<0.35$ ATLAS appears to perform better than redMaPPer in recovering Abell and ACT DR5 clusters\footnote{We note that the Abell sample is not as complete as the ATLAS and redMaPPer samples as the latter surveys have the advantage of detecting clusters in multiple colours, (with both catalogues detecting $\sim3-4$ times as many clusters as the Abell sample across their full survey footprints in the $z<0.3$ redshift range). However, Abell is still a useful sample for comparison to ATLAS and redMaPPer, as clusters detected in both ATLAS and Abell samples are likely to be genuine rich clusters which should be detected by redMaPPer.}. In the $0.35<z<0.55$ redshift range, however, redMaPPer appears to recover a larger fraction of ACT DR5 clusters than ATLAS. We also note that at higher redshifts there appears to be a larger number of redMaPPer clusters with no detections in ATLAS or ACT DR5 catalogues compared to lower redshift ATLAS clusters with no detections in the other catalogues. 

\begin{figure*}
	\begin{subfigure}[t]{\columnwidth}
		\centering
		\caption{}
		\includegraphics[width=0.8\columnwidth]{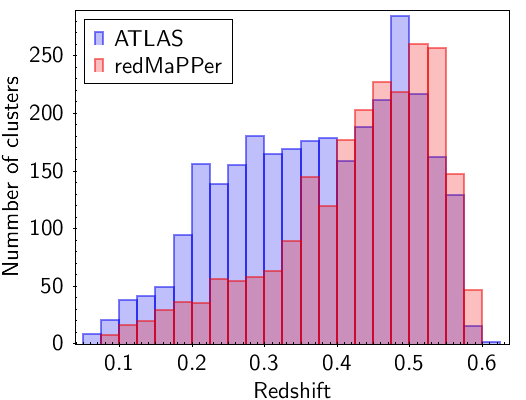}
		\label{fig:atlas_vs_redMaPPer_nz}
	\end{subfigure}
	\begin{subfigure}[t]{\columnwidth}
		\centering
		\caption{}
		\includegraphics[width=0.8\columnwidth]{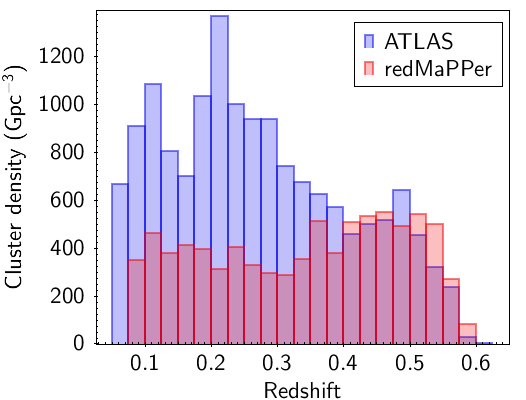}
		\label{fig:atlas_vs_redMaPPer_nz_density}
	\end{subfigure}
	\caption[Comparison of ATLAS and SDSS redMaPPer photometric redshift distributions]{(a) A comparison of the photometric redshift distribution of full ATLAS and SDSS redMaPPer cluster catalogues. Here, both samples are restricted to clusters with $M_{\textup{200m}}$>$3\times10^{14}h^{-1}$\(\textup{M}_\odot\). The redMaPPer histogram is scaled down by a factor of 2.2 to account for the difference between survey areas. (b) The number density of ATLAS and redMaPPer clusters per Gpc$^{3}$.}
	\label{fig:atlas_vs_redMaPPer_nzs}
\end{figure*}

Table~\ref{tab:ATLAS_RM_sky_density} shows a comparison of the sky density of the ATLAS cluster catalogue to the SDSS redMaPPer catalogue in five bins of redshift ranging from $z=0.05-0.55$. Similar to what we saw in Figure~\ref{fig:arm_extcat_map}, one can see here that in the redshift range $z<0.35$ the ATLAS catalogue has a $\sim 3\times$ higher cluster sky density compared to redMaPPer, while the two samples become more comparable at $z>0.35$. A similar pattern can be seen in Figure~\ref{fig:atlas_vs_redMaPPer_nz_density} where we compare the photometric redshift distributions of ATLAS and redMaPPer samples (without limiting  the two samples to the survey overlap areas).

\begin{table*}
	\centering
	\caption[Sky density of ATLAS clusters in comparison to the SDSS redMaPPer clusters as a function of redshift]{Number and sky density (number of clusters per square degree) of ATLAS and SDSS redMaPPer cluster samples in different redshift bins. Here, the samples are limited to the overlap area of the two surveys and we apply a mass cut of $M_{\textup{200m}}$>$3\times10^{14}h^{-1}$\(\textup{M}_\odot\) to both samples.}
	\label{tab:ATLAS_RM_sky_density}
	\begin{tabular}{c|cc|cc} 
		\hline
		 & \multicolumn{2}{c|}{Number of clusters} & \multicolumn{2}{c}{Sky density} \\
		  Redshift & ATLAS & redMaPPer & ATLAS & redMaPPer \\
		\hline
		$0.05<z<0.15$ & 7 & 2 & 0.03 & 0.01  \\
		$0.15<z<0.25$ & 21 & 7 & 0.10 & 0.03 \\
		$0.25<z<0.35$ & 47 & 14 & 0.22 & 0.07 \\
		$0.35<z<0.45$ & 56 & 52 & 0.26 & 0.24 \\
		$0.45<z<0.55$ & 90 & 72 & 0.42 & 0.34 \\
		\hline
	\end{tabular}
\end{table*}

To further examine this, we calculate the detection rate of Abell clusters (which are classified as clusters with 30 or more members), by the redMaPPer catalogue across the full SDSS survey footprint. We find that redMaPPer only recovers $\sim60\%$ of the $0.05<z<0.3$ Abell clusters, compared to a $\sim85\%$ recovery rate in the ATLAS sample with a mass cut of $M_{\textup{200m}}$>$0.9\times10^{14}h^{-1}$\(\textup{M}_\odot\) (which approximately corresponds to the lower mass limit of the redMaPPer sample given its $\lambda>20$ richness cut). Consequently, the higher number of ATLAS detections in this redshift range relative to redMaPPer is likely to be predominantly due to the incompleteness of the redMaPPer sample at lower redshifts.

We now explore the completeness and purity of the ATLAS sample as a function of redshift, based on direct comparison to redMaPPer in the $M_{\textup{200m}}$>$3\times10^{14}h^{-1}$\(\textup{M}_\odot\) mass range. In column (2) of Table~\ref{tab:ARM_comp_pur}, for each redshift bin, we calculate the fraction of SDSS redMaPPer clusters (within the ATLAS coverage area) that are also detected by {\sc orca} using the ATLAS data\footnote{Similar to Section~\ref{sec:Mass_redshift_completeness}, we use the {\sc orca} radius $R_{\rm ORCA}$ as our matching radius here and throughout the rest of this Section when comparing the two catalogues.}. Here, our aim is to simply compute the likelihood that a SDSS redMaPPer cluster detection is also identified by {\sc orca} in the ATLAS data. As such, when performing the matching, we only impose the mass and redshift cuts on the redMaPPer catalogue to allow for matches to be made in cases where the two catalogues assign the same cluster to different mass and redshift bins. These selections in effect ensure that our completeness estimates are not biased due to differences in photometric redshift and richness estimates between the two samples.  

With these caveats in mind, we convert the recovery fractions shown in column (2) to percentages giving an estimate of the completeness of the ATLAS catalogue as a function of redshift (under the assumption that the redMaPPer sample is $100\%$ pure). Based on this comparison, we find that the ATLAS sample is $100\%$ complete up to $z<0.25$, with a slight reduction in completeness to $93\%$ between $0.25<z<0.35$ and a further reduction down to $63\%$ and $46\%$ completeness at $0.35<z<0.45$ and $0.45<z<0.55$ respectively. Although these values are presented here to show a full comparison of the two samples, we note that the mean completeness shown in Figure~\ref{fig:Mean_redshift_completeness} is likely to be a more reliable estimate of the completeness of the ATLAS cluster sample than estimates based on comparison to any one external sample alone.

\begin{table*}
	\centering
	\caption[Statistical comparisons of ATLAS and SDSS redMaPPer cluster detection rates]{Various comparisons of the intersections between the ATLAS and SDSS redMaPPer cluster catalogues. (columns 2 \& 3) Fraction \& percentage of SDSS redMaPPer clusters with detections in the ATLAS cluster catalogue. (4) The fraction of ATLAS clusters with counterpart detections in the redMaPPer catalogue. (5) Among the ATLAS clusters not matched to redMaPPer in column (4), how many clusters we constitute as genuine and rich detection clusters based on Visual Inspection (VI) of their photo-z $n(z)$. (6) The fraction of ATLAS detections that are either detected by redMaPPer or appear as likely detections based on the VI of the cluster $n(z)$s. (7) An estimate of the completeness of the redMaPPer  based on the ratio of redMaPPer cluster detections to the number of "pure" ATLAS detections (from column 6). We refer the reader to the discussion in the text for full details of the sample selections implemented prior to comparing the catalogues.}
	\label{tab:ARM_comp_pur}
	\begin{tabular}{ccc|ccc|c} 
		\hline
		 Redshift & AT/RM & AT Comp. & RM/AT & VI & AT Purity & RM Comp.\\
		 (1) & (2) & (3) & (4) & (5) & (6) & (7)\\
		\hline
		$0.05<z<0.15$ & 2/2  & 100\% & 5/7 & +2/7 & 7/7 (100\%) & 5/7 (71\%) \\[0.1cm] 
		$0.15<z<0.25$ & 7/7  & 100\% & 16/21 & +6/21 & 21/21 (100\%) & 16/21 (76\%) \\[0.1cm] 
		$0.25<z<0.35$ & 13/14 & 93\% & 29/47 & +12/47 & 41/47 (87\%) & 29/41 (70\%) \\[0.1cm] 
		$0.35<z<0.45$ & 35/52 & 63\% & 30/56 & +10/56 & 40/56 (71\%) & 30/40 (75\%) \\[0.1cm] 
		$0.45<z<0.55$ & 33/72 & 46\% & 23/90 & +27/90 & 50/90 (56\%) & 23/50 (46\%) \\[0.1cm]
		\hline
	\end{tabular}
\end{table*}

In columns (4, 5 \& 6) of Table~\ref{tab:ARM_comp_pur}, we estimate the purity of the ATLAS cluster catalogue based on comparison to the cluster detections by redMaPPer in the survey overlap region. As we are interested in assessing our sample purity as a function of redshift, we limit the ATLAS sample to photometric redshift bins shown in column (1) and look for confirmation that the ATLAS detection is also identified by redMaPPer. The fraction of ATLAS detections confirmed by redMaPPer in this way at each redshift bin is shown in column (4). 

\begin{figure*}
	\begin{subfigure}[c]{\columnwidth}
		\centering
		\caption{}
		\includegraphics[width=0.7\columnwidth]{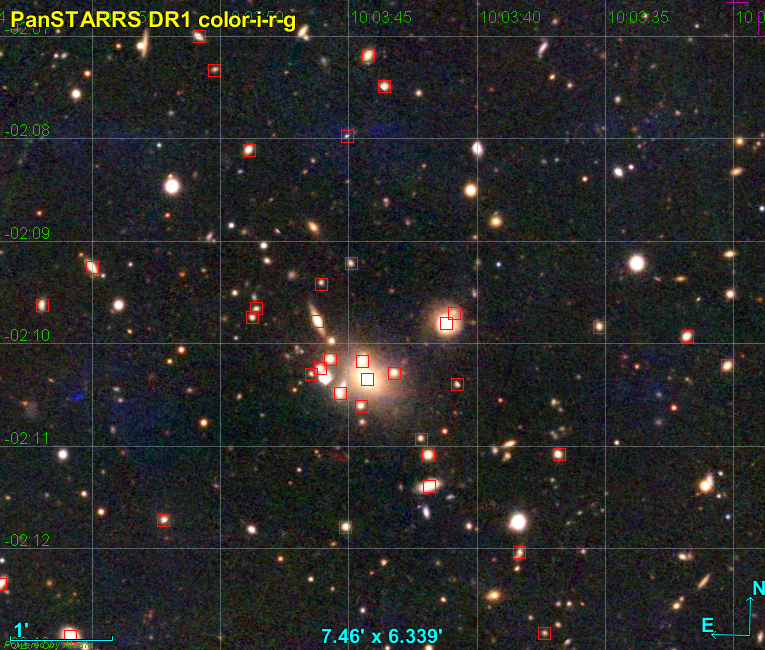}
		\label{
		fig:AT_11_v2}
	\end{subfigure}	
	\begin{subfigure}[c]{\columnwidth}
	    \centering
		\caption{}
		\includegraphics[width=0.7\columnwidth]{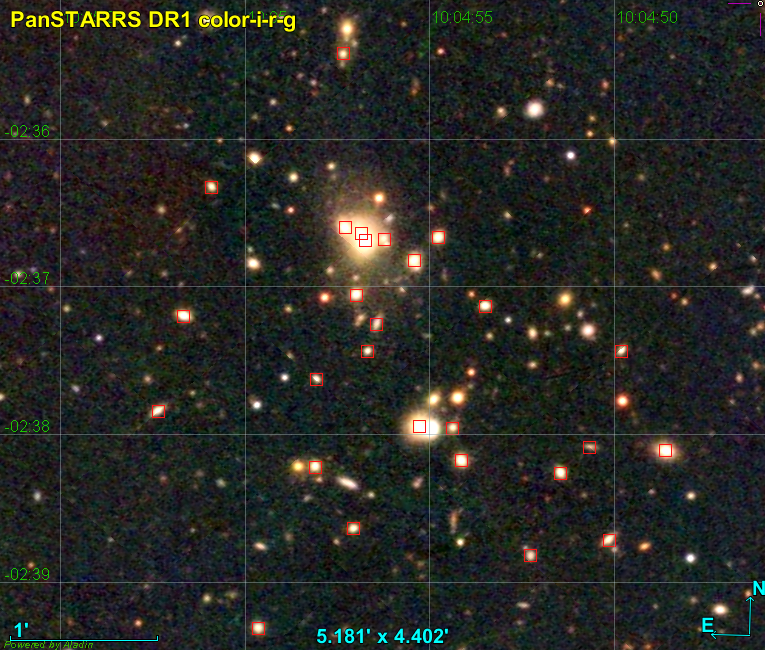}
		\label{fig:AT_15_v2}
	\end{subfigure}
	\begin{subfigure}[c]{\columnwidth}
		\centering
		\caption{}
		\includegraphics[width=0.6\columnwidth]{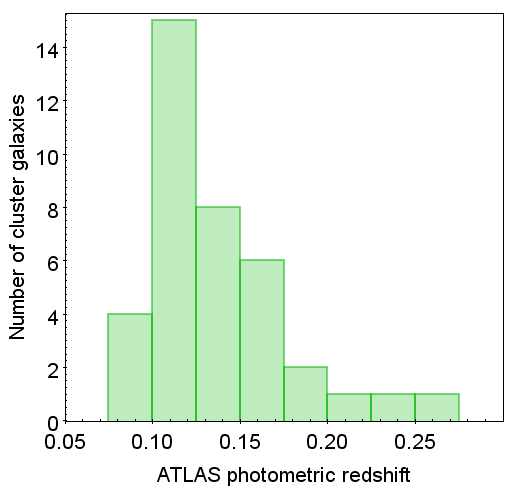}
		\label{fig:AT_11_hist}
	\end{subfigure}
	\begin{subfigure}[c]{\columnwidth}
		\centering
		\caption{}
		\includegraphics[width=0.6\columnwidth]{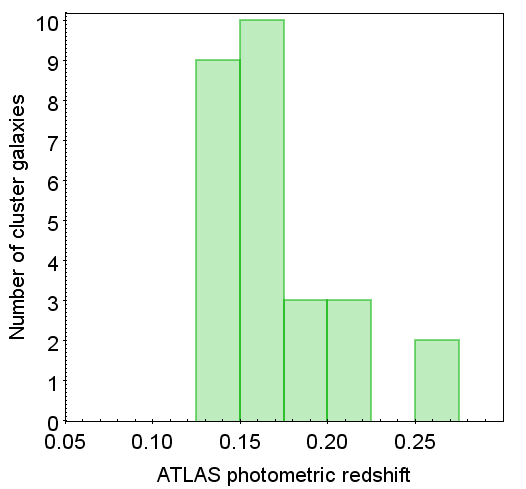}
		\label{fig:AT_15_hist}
	\end{subfigure}
	\caption[Visually confirmed ATLAS clusters not detected in SDSS by redMaPPer.]{(a \& b) PanSTARRS colour images of two $N_{200}>20$ ATLAS clusters with no redMaPPer detections in SDSS. Here the cluster members identified in ORCA are shown by the red squares. (c \& d) The photometric redshift histograms for clusters in panels (a \& b) respectively.}
	\label{fig:ATLAS_not_rm}
\end{figure*}

We note, however, that without access to the full redMaPPer catalogue we have no way of checking whether our clusters with no successful matches to redMaPPer may have been detected and classified by redMaPPer as clusters with $\lambda<20$, or indeed, remain undetected by redMaPPer. To overcome this limitation, in the next step we visually inspect the photometric redshifts of the clusters that were detected as $M_{\textup{200m}}$>$3\times10^{14}h^{-1}$\(\textup{M}_\odot\) systems in ATLAS, but did not have a successful match in redMaPPer. We also count our cluster detections as "pure" (i.e. non-spurious detections) if $>50\%$ of the cluster members appear concentrated in a histogram with a width of $\bar{z}\pm0.025$ (corresponding to the ATLAS RMS error on cluster galaxy photometric redshifts). The choice of this criterion was motivated based on visual inspection of the photometric redshift histograms of ATLAS clusters with successful matches to redMaPPer, where almost all clusters detected in both catalogues had ATLAS or SDSS photometric redshift histograms which were centred around a redshift with more than half of cluster members lying within $\bar{z}\pm0.025$ of the histogram peak. Figure~\ref{fig:ATLAS_not_rm} shows colour images and photometric redshift histograms of two $M_{\textup{200m}}$>$3\times10^{14}h^{-1}$\(\textup{M}_\odot\) ATLAS clusters with no SDSS redMaPPer detections, which were confirmed as pure following the procedure described above.

The fraction of additional clusters confirmed as genuine cluster detections based on visual inspection is presented in column (5), with column (6) showing the sum of the fractions in columns (4 \& 5) with the percentages in this column providing an estimate of the purity of the ATLAS cluster detections. Based on this estimate we find our $M_{\textup{200m}}$>$3\times10^{14}h^{-1}$\(\textup{M}_\odot\) clusters to be $100\%$ pure at $z<0.25$, and $87\%$ pure in the redshift range $0.25<z<0.35$. The purity of the sample then falls to $71\%$ and $56\%$ in our final two redshift bins. This reduction of sample purity with increasing redshift could be partially due to the fact that as we reach the magnitude limits of the ATLAS and SDSS surveys the likelihood of real but faint cluster members not being detected by redMaPPer increases, bringing clusters with richness greater than 20, below their $\lambda>20$ limit. Similarly, the increase in the  uncertainty on our estimated photometric redshifts with increasing redshift, reduces the number of clusters classified as pure on the basis of their photo-z histograms in column (5). On the other hand, it is also possible that at higher redshifts redMaPPer performs better than {\sc ORCA} in overcoming projection effects which could artificially increase the richness of our clusters, or result in false cluster detections. We refer the reader to Section 6.4.3 of \cite{Murphy2012}, for a different analysis of the purity of {\sc orca} cluster detections as a function of cluster mass and redshift, based on comparison to SDSS-like simulated mock cluster catalogues. Based on this comparison, the purity of {\sc orca} detections at the median redshift of the survey was shown to be $>70\%$ across all cluster masses.

Column (7) of Table~\ref{tab:ARM_comp_pur} shows an estimate of the completeness of the redMaPPer sample, which is defined as the fraction of "pure" ATLAS detections co-detected by redMaPPer\footnote{We refer the reader to Section 11 of \cite{Rykoff2014} for a more systematic analysis of the completeness of the redMaPPer sample.}. We find the completeness of redMaPPer to be in the $\sim70-75\%$ range at $z<0.45$, before falling down to $\sim50\%$ at $0.45<z<0.55$. At this point we remind the reader that although we were able to estimate the completeness of the ATLAS sample in column (3) by including successful matches between ATLAS and redMaPPer in cases where a $\lambda>20$ redMaPPer cluster had an ATLAS detection with $N_{200}<20$, we are unable to do the reverse in column 6. Here we recall that, while it is possible that some of our ATLAS detections may have redMaPPer counterparts with $\lambda<20$ which would increase the redMaPPer completeness estimate; we have no way of verifying this. Indeed, this could explain the apparent fall in the completeness of the redMaPPer sample in the highest redshift bin where it is more likely for cluster richness to be under-estimated due to the increased likelihood of cluster galaxies falling out of the survey magnitude limits. Nonetheless, the information in column (7) provides a useful estimate of the percentage of low-redshift ATLAS clusters with $M_{\textup{200m}}$>$3\times10^{14}h^{-1}$\(\textup{M}_\odot\) that are missed by redMaPPer, (possibly due to differences in our definitions of cluster richness and what constitutes a "pure" cluster detection).

To summarise, comparison of the ATLAS cluster catalogue to redMaPPer shows that at $z<0.35$ the ATLAS sample is highly complete, and the majority of our clusters appear to be genuine detections. At $z<0.35$ we also find that a number of $M_{\textup{200m}}$>$3\times10^{14}h^{-1}$\(\textup{M}_\odot\) ATLAS clusters (which were visually confirmed as rich clusters) are not detected as $\lambda>20$ clusters by redMaPPer. Similarly, we found that ATLAS generally performs better than redMaPPer at recovering $z<0.35$ Abell and ACT DR5 clusters, with ATLAS recovering $\sim85\%$ of $0.05<z<0.3$ Abell clusters while redMaPPer only recovers $\sim60\%$ of Abell clusters. At $z>0.35$ while ATLAS appears to be less complete ($65\%$) compared to redMaPPer, Figure~\ref{fig:atlas_vs_redMaPPer_nz_density} suggests that the additional redMaPPer cluster detections at higher redshifts tend to be clusters of lower mass with a good agreement being found between the $n(z)$ of the two samples for clusters with masses $M_{\textup{200m}}$>$3\times10^{14}h^{-1}$\(\textup{M}_\odot\).

\subsection{ATLAS cluster masses}
\label{sec:ORCA_masses_results}

\begin{figure*}
	\begin{subfigure}[T]{\columnwidth}
		\centering
		\subcaption{}
		\vspace{-0.2\baselineskip}
		\includegraphics[width=\columnwidth]{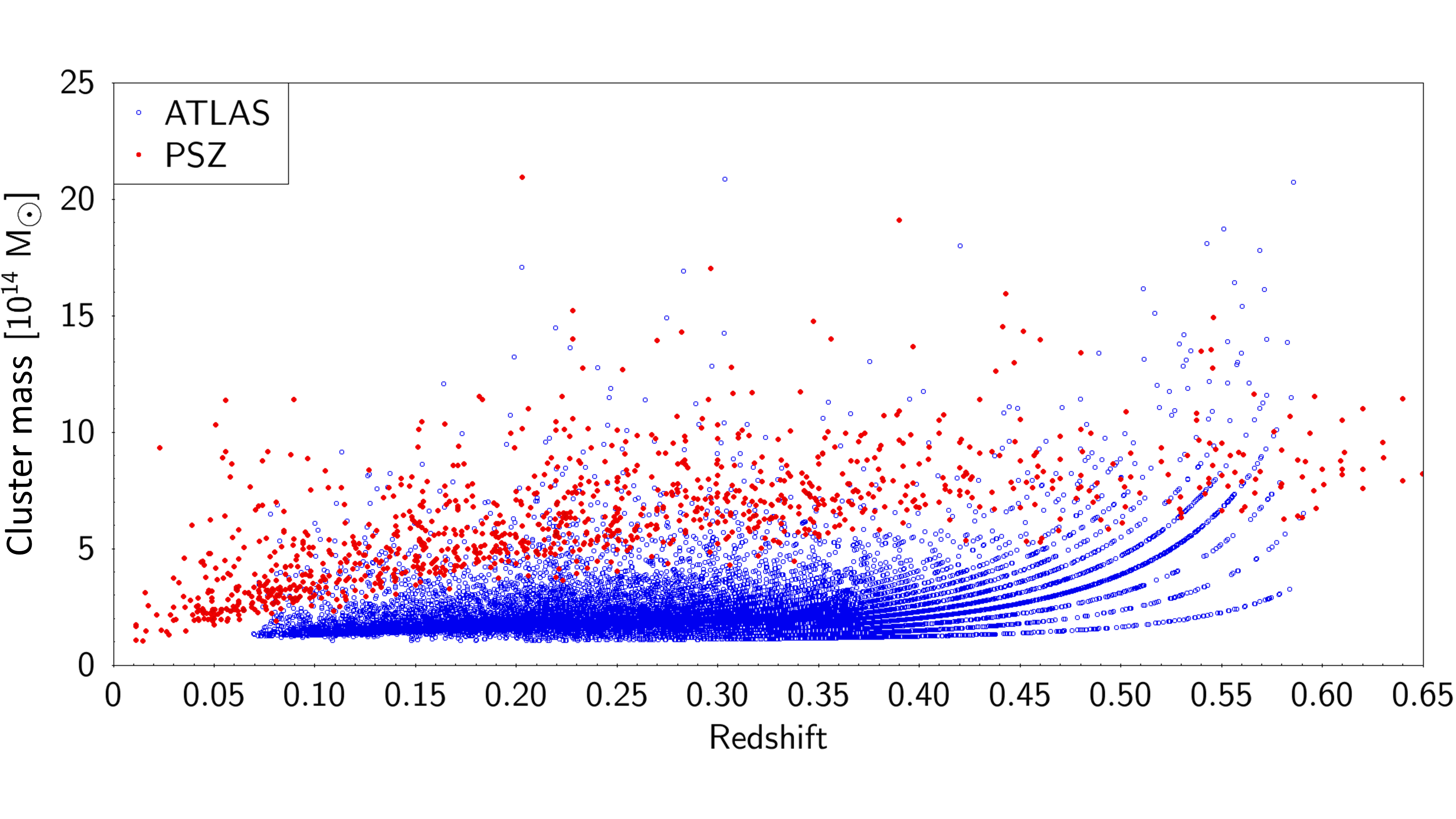}
		\label{fig:ATLAS+PSZ}
	\end{subfigure}
	\vspace{-1.5\baselineskip}
	\begin{subfigure}[T]{\columnwidth}
		\centering
		\subcaption{}
		\includegraphics[width=\columnwidth]{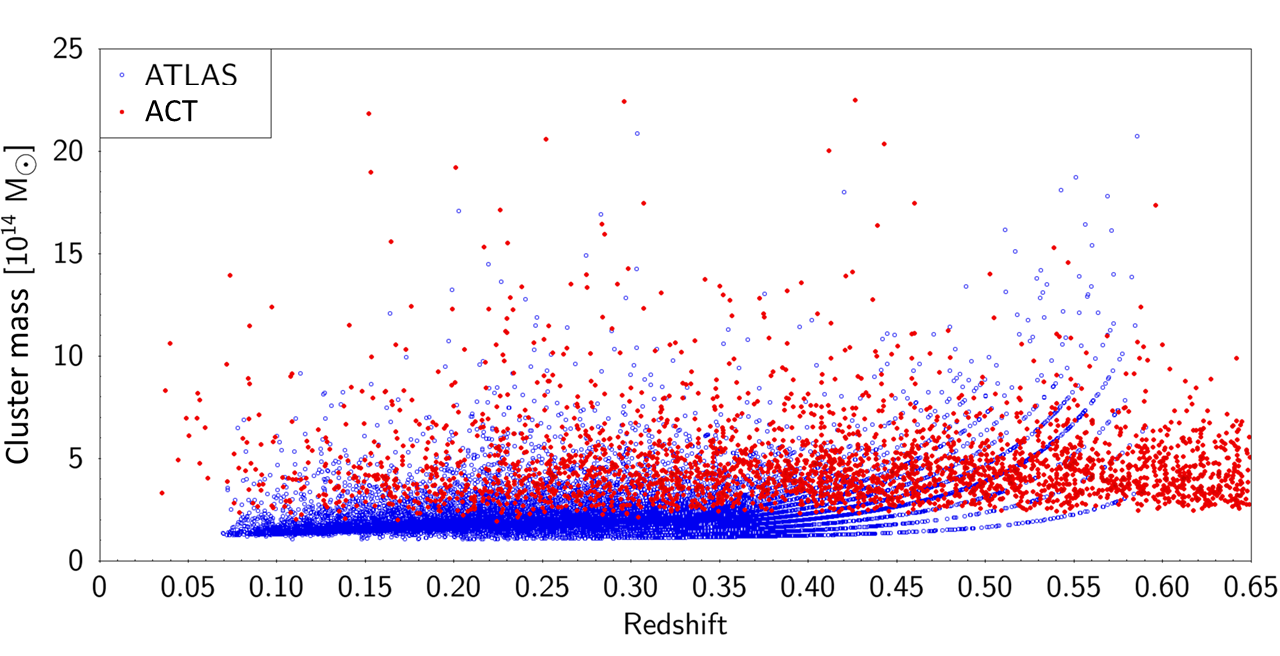}
		\label{fig:ATLAS+ACT}
	\end{subfigure}
	\vspace{-1.5\baselineskip}
	\begin{subfigure}[T]{\columnwidth}
		\centering
		\caption{}
		\vspace{-0.2\baselineskip}
		\includegraphics[width=\columnwidth]{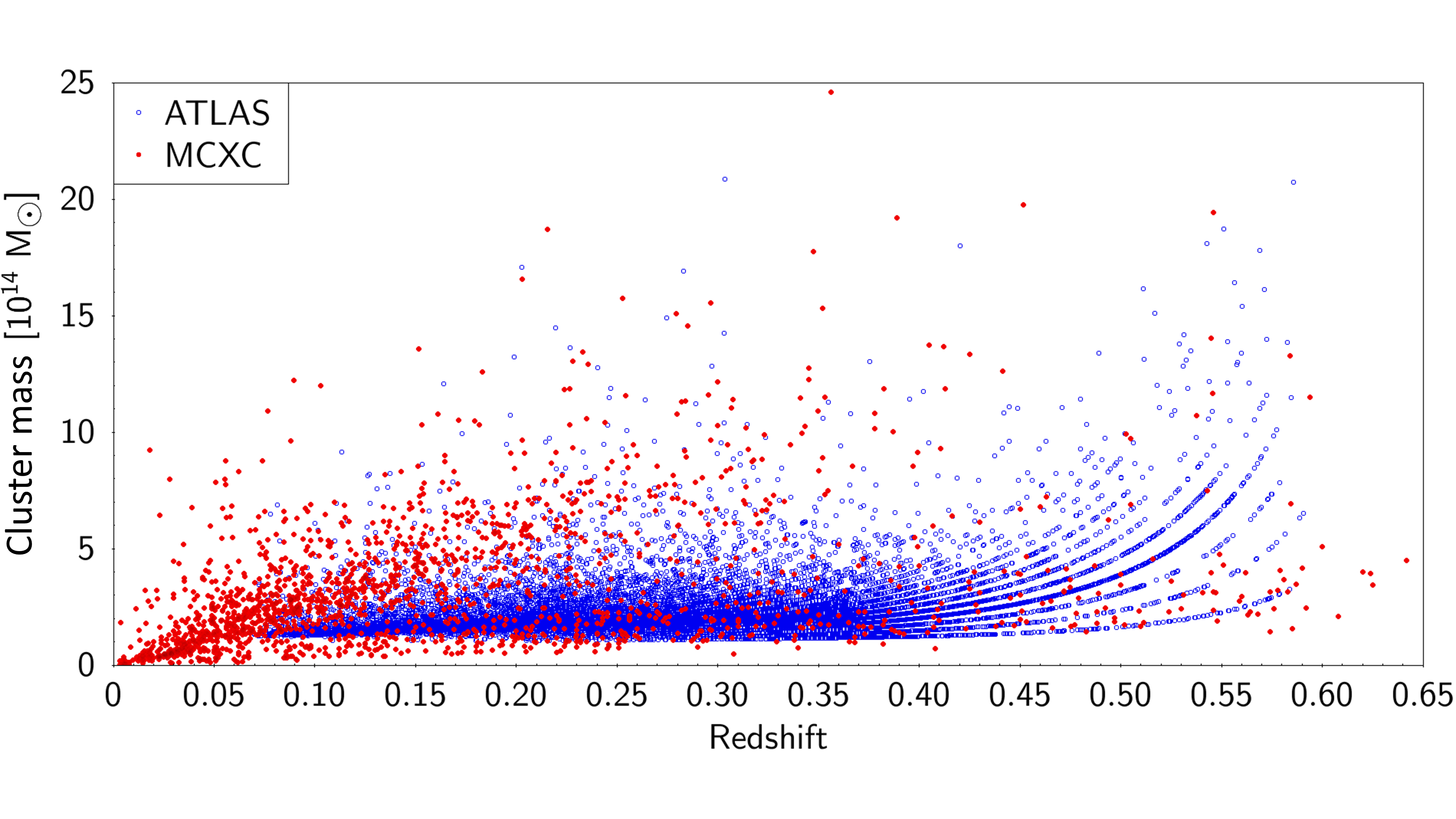}
		\label{fig:ATLAS+MCXC}
	\end{subfigure}
	\begin{subfigure}[T]{\columnwidth}
		\centering
		\caption{}
		\includegraphics[width=\columnwidth]{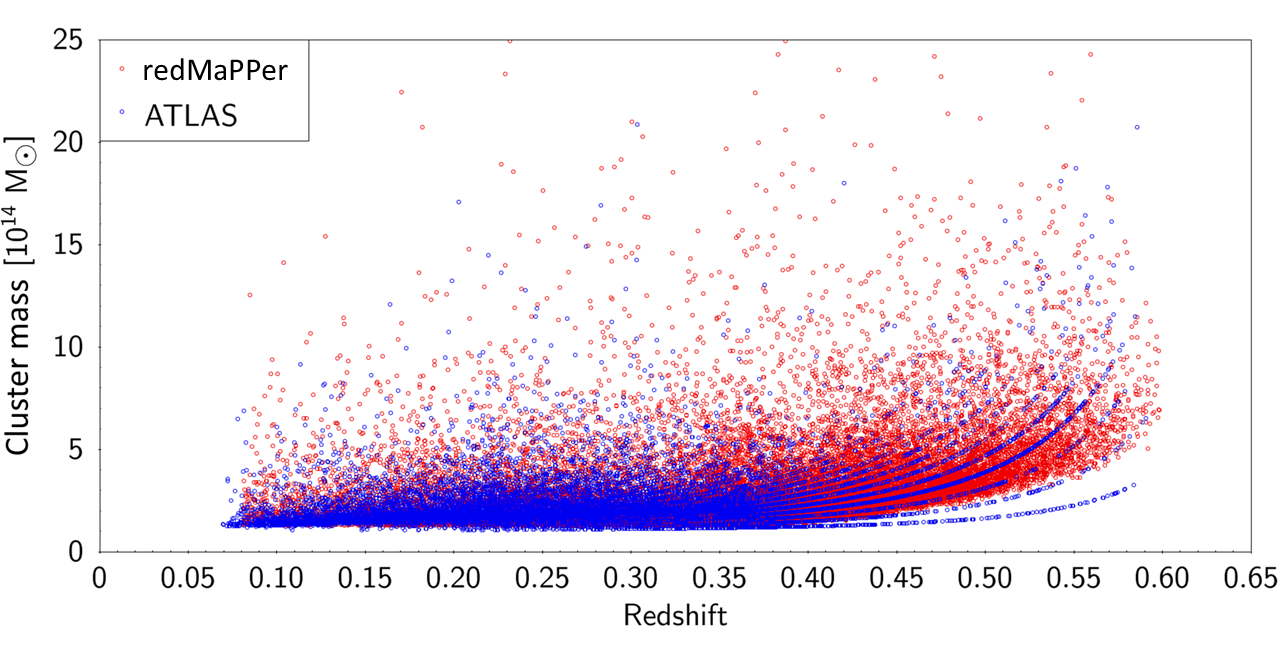}
		\label{fig:ATLAS+redMaPPer}
	\end{subfigure}
	\caption[Comparison of ATLAS cluster masses to ACT DR5 and Planck SZ cluster masses]{Comparison of the ATLAS $M_{\textup{200m}}$ cluster masses to Planck SZ masses $M_{\textup{SZ}}$ (panel a); ACT DR5 $M_{\textup{200m}}$ SZ masses (panel b); MCXC $M_{500}$ cluster masses (panel c), and SDSS redMaPPer $M_{200m}$ cluster masses (panel d).}
	\label{fig:ORCA_vs_ACT_PSZ}
\end{figure*}

Figure~\ref{fig:ORCA_vs_ACT_PSZ} shows a comparison of the ATLAS cluster masses, obtained using Equation~\ref{eq:M_200}, to SZ cluster masses of the full ACT DR5 and Planck samples. As seen in Figure~\ref{fig:ATLAS+PSZ} the ATLAS cluster catalogue provides a complementary sample to the Planck catalogue, by detecting clusters in a lower mass range than possible with Planck. On the other hand, SZ cluster surveys offer the ability to detect all clusters above a certain mass threshold with little dependence on redshift, whereas optical surveys are limited in their ability to detect clusters at higher redshifts by the magnitude limit of the survey. As a result, the two approaches to cluster detection are highly complementary in maximizing the mass and redshift completeness of cluster samples.

As seen in Figure~\ref{fig:ATLAS+ACT}, the ATLAS cluster sample also complements the ACT DR5 SZ sample in terms of detection of lower mass clusters in the redshift range $z<0.5$. We note however, that due to it's higher angular resolution and superior flux sensitivity, ACT DR5 is able to detect lower mass SZ clusters than Planck. The larger Planck beam size, however, makes it more sensitive to clusters at $z<0.05$ when compared to ACT DR5 (due to the larger projected area of these low redshift clusters on the sky).

Figure~\ref{fig:ATLAS+MCXC} shows a comparison of our ATLAS cluster masses, with X-ray cluster masses ($M_{500}$) from the MCXC sample. While most X-ray surveys included in the MCXC sample tend to be more sensitive to low-mass, low-redshift clusters in comparison to optical cluster catalogues, the completeness of these X-ray flux-limited cluster samples is reduced with redshift, with optical and SZ surveys providing more complete cluster samples at higher redshifts. Nevertheless, X-ray cluster samples play an important role in calibrating optical and SZ cluster masses and in determining the cluster gas fractions. The next generation of deeper X-ray cluster samples such as eROSITA will significantly improve on the completeness and statistics of current X-ray cluster samples by detecting $\sim100,000$ galaxy clusters with masses $>5\times10^{13} h^{-1}$\(\textup{M}_\odot\) \citep{Pillepich2012}, a 2 orders of magnitude improvement over the MCXC cluster sample size. 

Finally, we compare the ATLAS cluster masses to cluster masses from the SDSS redMaPPer cluster catalogue in Figure~\ref{fig:ATLAS+redMaPPer}. It can be seen that the two samples cover a similar range of cluster masses and redshifts, which is expected, given that both samples detect clusters using optical $griz$ bands with a similar depth. However, as the ATLAS catalogue covers areas of the southern sky, not covered by the SDSS redMaPPer catalogue, the two catalogues are highly complementary. For reasons which we discussed in Section~\ref{sec:RM_comparison}, the ATLAS sample also contains a larger number of cluster detections in the redshift range $0.05<z<0.3$, while the redMaPPer catalogue contains more cluster detections in the $0.3<z<0.55$ range, making this another aspect in which the two catalogues are complementary. 

\subsection{Comparison of cluster mass functions}
\label{sec:arm_cluster_mfs}

\begin{figure*}
	\begin{subfigure}[T]{\columnwidth}
		\centering
		\caption{}
		\vspace{-0.2\baselineskip}
		\includegraphics[width=\columnwidth]{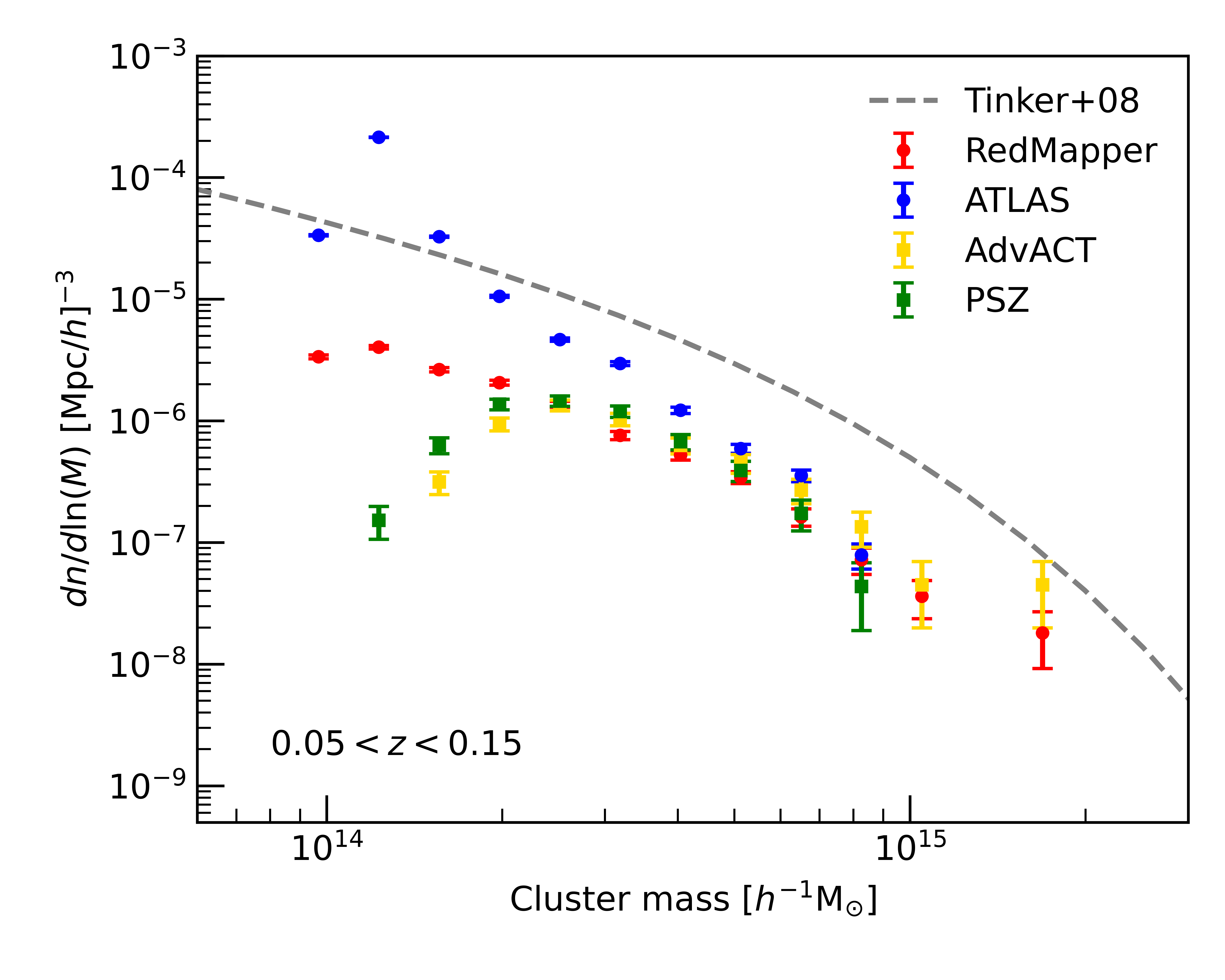}
		\label{fig:psz_sep_z0p1}
	\end{subfigure}\vspace{-0.5\baselineskip}	
	\begin{subfigure}[T]{\columnwidth}
		\centering
		\caption{}
		\vspace{-0.2\baselineskip}
		\includegraphics[width=\columnwidth]{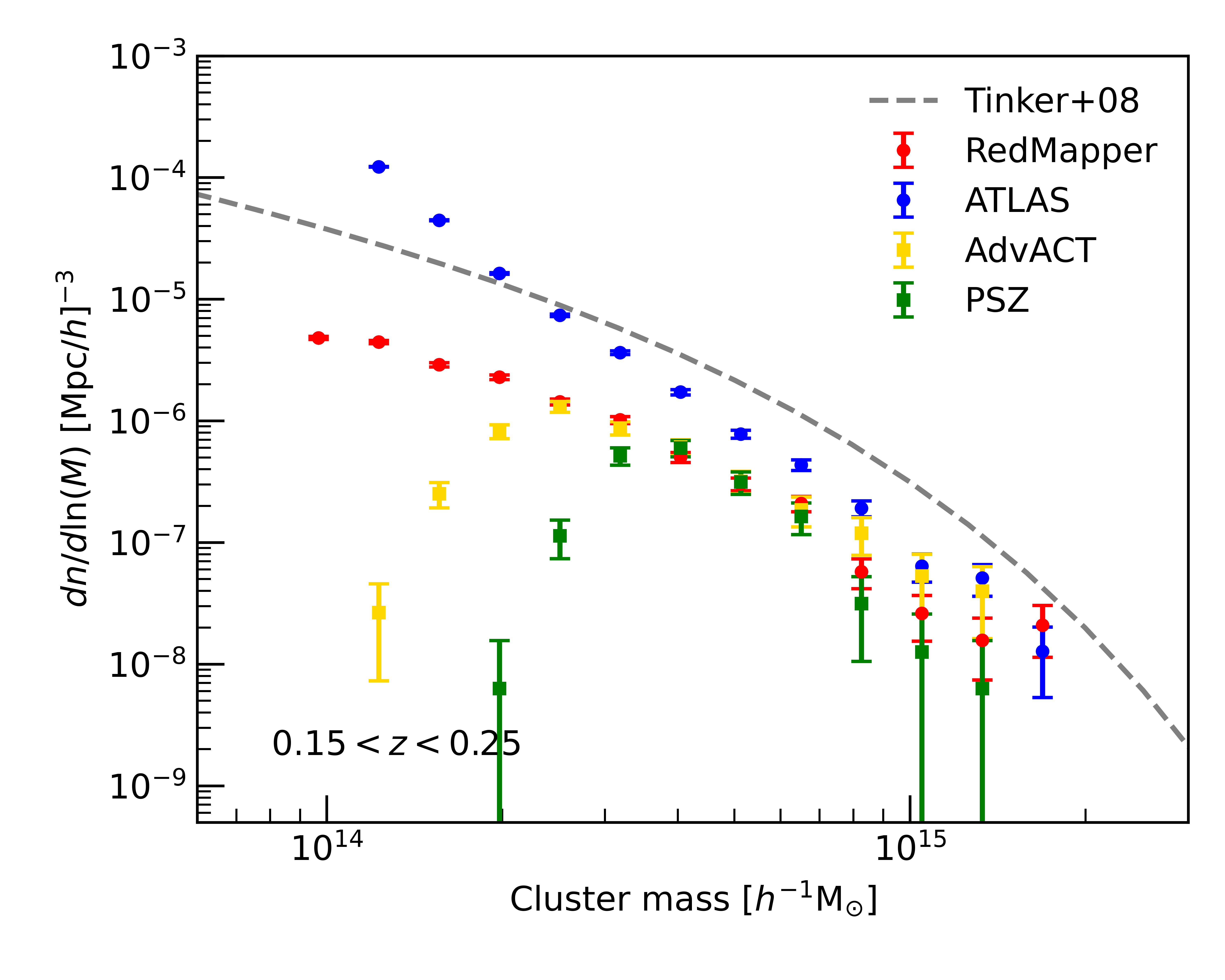}
		\label{fig:psz_sep_z0p2}
	\end{subfigure}\vspace{-0.5\baselineskip}
	\begin{subfigure}[T]{\columnwidth}
		\centering
		\caption{}
		\vspace{-0.2\baselineskip}
		\includegraphics[width=\columnwidth]{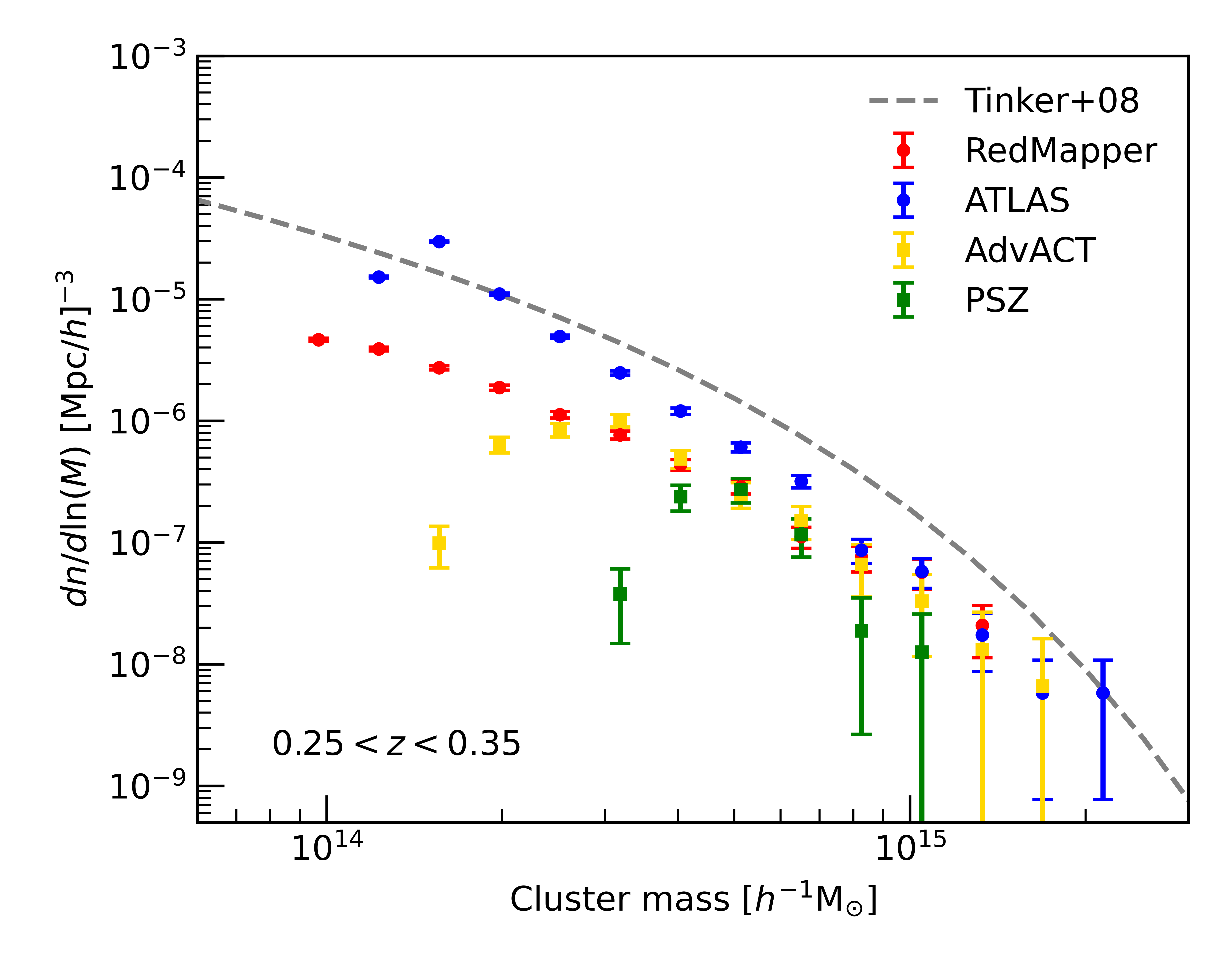}
		\label{fig:psz_sep_z0p3}
	\end{subfigure}\vspace{-0.5\baselineskip}
	\begin{subfigure}[T]{\columnwidth}
		\centering
		\caption{}
		\vspace{-0.2\baselineskip}
		\includegraphics[width=\columnwidth]{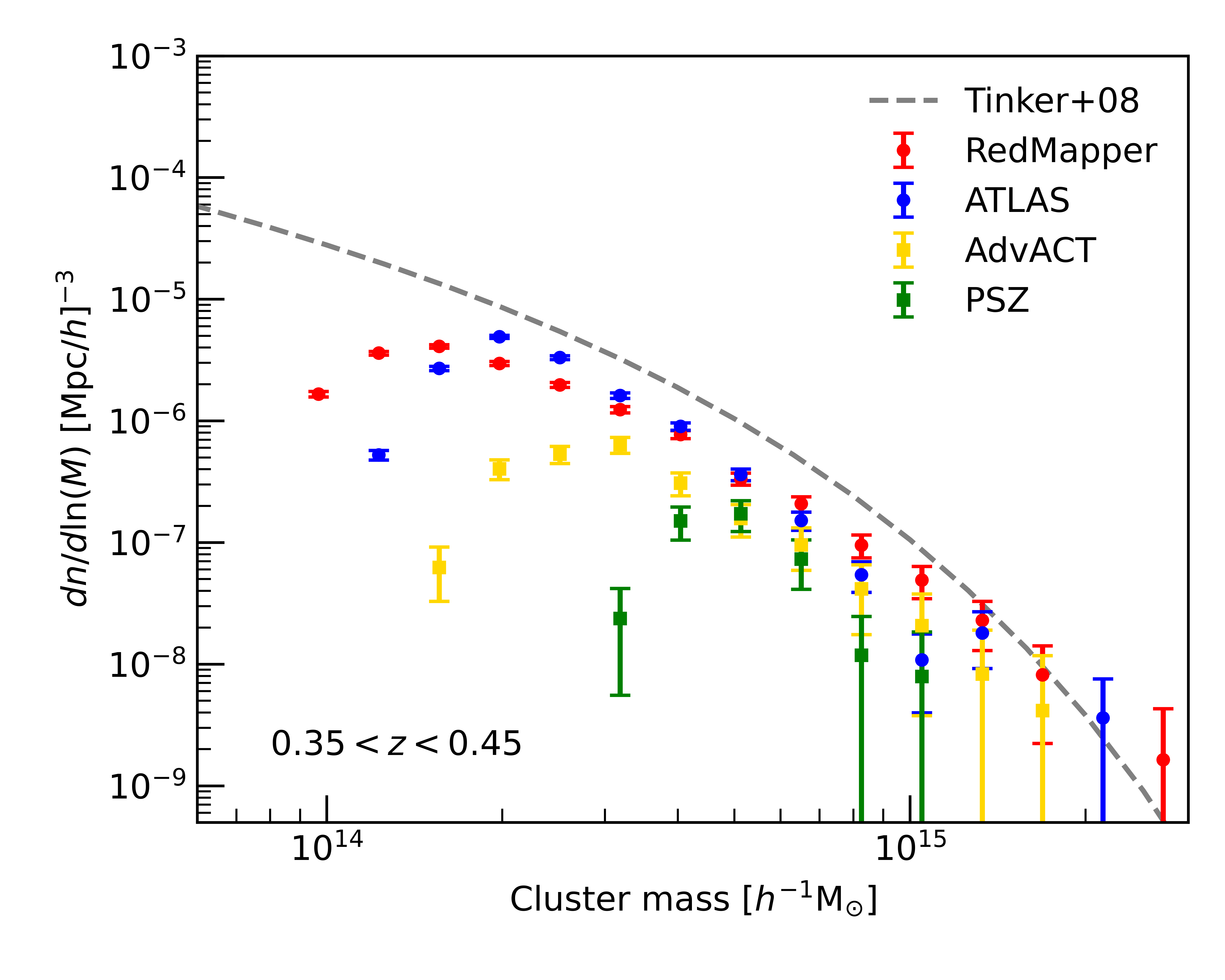}
		\label{fig:psz_sep_z0p4}
	\end{subfigure}\vspace{-0.5\baselineskip}
	\begin{subfigure}[T]{\columnwidth}
		\centering
		\caption{}
		\vspace{-0.2\baselineskip}
		\includegraphics[width=\columnwidth]{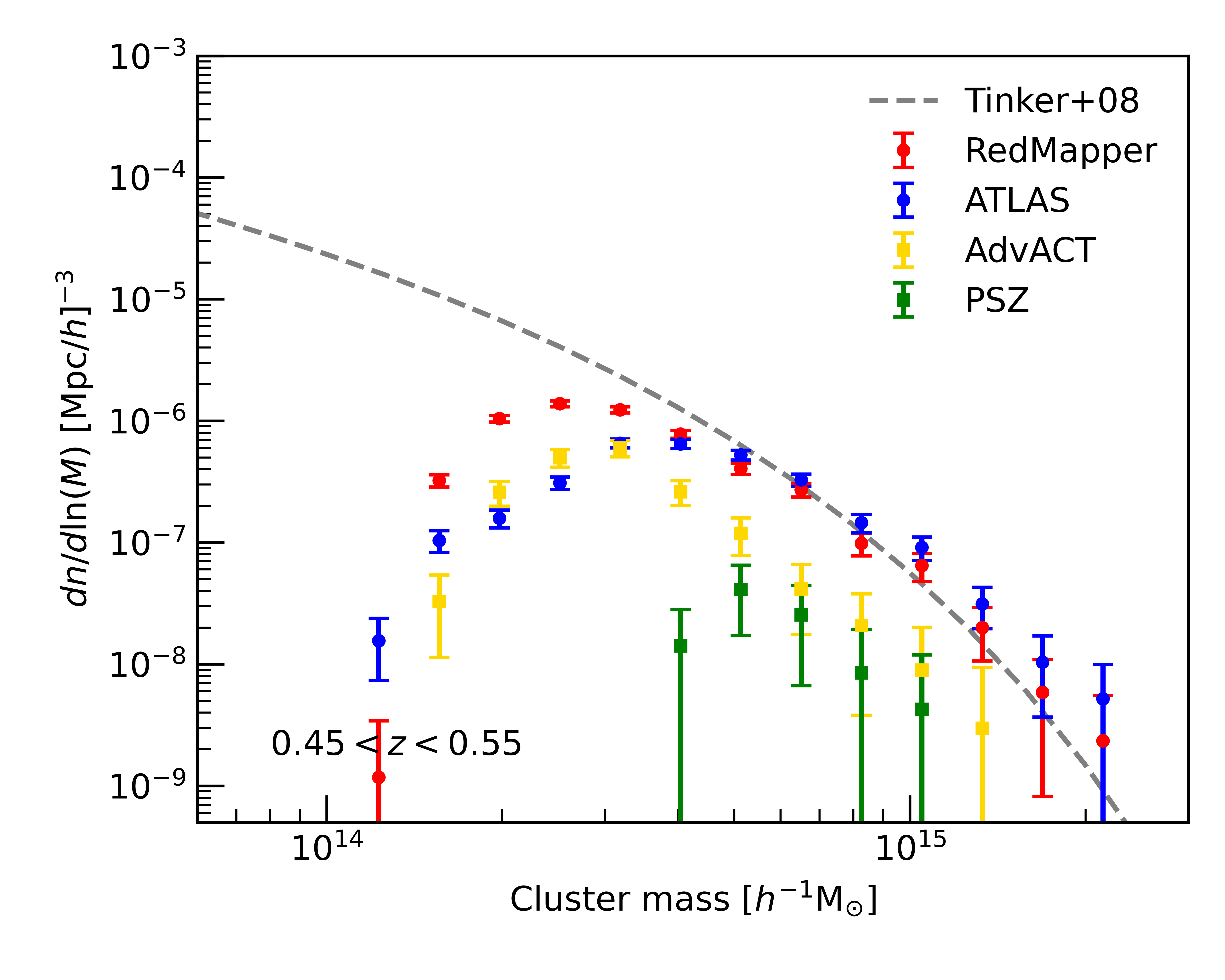}
		\label{fig:psz_sep_z0p5}
	\end{subfigure}\vspace{-0.5\baselineskip}
	\caption[Redshift evolution of optical and SZ cluster mass functions]{A comparison of ATLAS, redMaPPer, Planck and ACT DR5 cluster mass functions to the theoretical predictions of the \cite{Tinker2008} $\Lambda$CDM models assuming a \cite{Planck2016XIII} cosmology. Here, the error bars on the observed mass functions are simply estimated as $\sqrt{n}$.}
	\label{fig:SZ_mfs}
\end{figure*}

We now compare the observed cluster mass functions of ATLAS, redMaPPer, Planck and ACT DR5 samples to the theoretical predictions of the \cite{Tinker2008} model. The mass function models are generated for spherical overdensities defined as $M_{200m}$, assuming the Planck 2015 cosmology (\citealt{Planck2016XIII}; Table 4, column 6). We compare the observed mass functions to the models in five redshift bins, in order to examine the evolution of the cluster mass function in the redshift range $0.05<z<0.55$.

In Figure~\ref{fig:SZ_mfs} we compare the redshift evolution of observed optical and SZ cluster mass functions to the predictions of $\Lambda$CDM. For the purpose of visual comparison of the cluster mass functions, the error bars shown in this plot are simply estimated by $\sqrt{n}$, where $n$ is the number of clusters in each mass bin. However, in Paper II, we shall perform a more detailed analysis of the mass function uncertainties prior to using the mass functions for cosmological parameter extraction. The fall in the amplitude of the cluster mass functions at lower masses seen in this figure is due to the reduced completeness of the samples as they approach their lower limit of cluster mass detection. As expected given the selection functions of ACT DR5 and Planck, the optical samples probe a lower range of cluster masses than the two SZ samples. 

As seen in Figure~\ref{fig:SZ_mfs} in the $0.05<z<0.35$ redshift range, we find the ATLAS cluster mass functions to have a higher amplitude relative to redMaPPer, placing them in better agreement with the $\Lambda$CDM model predictions. Despite this closer agreement however, strong and unexplained tensions between the observed ATLAS mass functions and the model predictions are present, particularly at higher masses. The lower amplitude of redMaPPer could be in part due to an underestimation of redMaPPer cluster masses at lower redshifts. This is in line with the discussion presented in Section VI.B of \cite{DES2020}, where the lower than expected $\Omega_\textup{m}$ obtained from cluster counts of the DES redMaPPer sample is attributed to a possible underestimation of the weak lensing estimated cluster masses for $\lambda<30$ redMaPPer clusters.

However, in Section~\ref{sec:RM_comparison} we also found a higher density (per unit volume) of ATLAS clusters at $z<0.35$ compared to redMaPPer. Similarly, our comparison of ATLAS and redMaPPer samples to the Abell cluster catalogue at lower redshifts showed that for clusters with $N_{200}>20$, ATLAS recovers $\sim85\%$ of Abell clusters, while redMaPPer recovers Abell clusters at a lower rate of $\sim60\%$. This could be taken as an indication that at lower redshifts redMaPPer is likely to miss a larger fraction of genuine (and relatively rich) clusters compared to ATLAS. Consequently, redMaPPer's lower completeness at $z<0.35$ could also be an important contributing factor towards the lower amplitudes of its mass function relative to $\Lambda$CDM predictions. 

As one can see in Figure~\ref{fig:SZ_mfs}, although the ACT DR5 survey probes a lower range of cluster masses than Planck\footnote{Note that in the $z=0.1$ bin (Figure~\ref{fig:psz_sep_z0p1}) the fact that Planck probes lower redshifts than ACT DR5, offsets ACT DR5's ability to probe clusters of lower masses, resulting in both samples having a similar completeness in the $M_{\textup{200m}}$<$3\times10^{14}h^{-1}$\(\textup{M}_\odot\) mass range.}, at all redshifts we find a good agreement between the mass functions of the two SZ samples in the range of masses where both surveys are complete. With the exception of the $z=0.5$ redshift bin (Figure~\ref{fig:psz_sep_z0p5}), the Planck and ACT DR5 mass functions also appear to be in general agreement with redMaPPer, placing them below the ATLAS cluster mass functions and the predictions of $\Lambda$CDM.  

\begin{figure*}
	\begin{subfigure}[T]{\textwidth}
		\centering
		\includegraphics[width=0.9\textwidth]{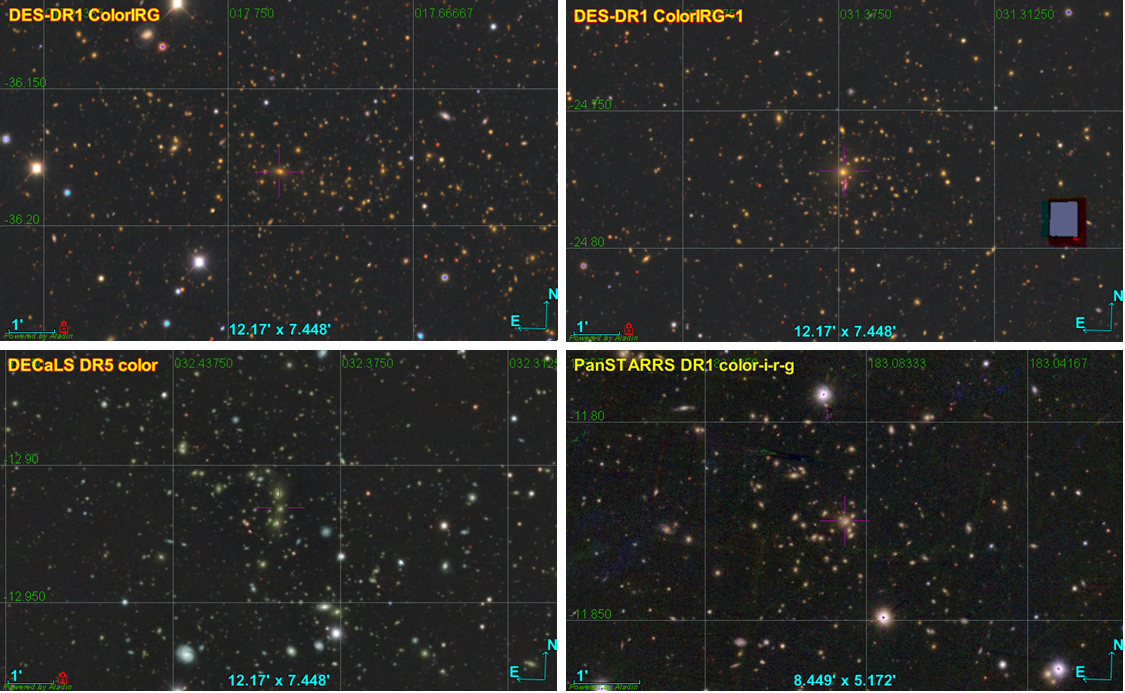}
	\end{subfigure}
	\caption[Examples of four rich Abell and ATLAS clusters not detected in the Planck SZ cluster samples]{DES, DECaLS or PanSTARRS images of four $z\sim0.2$, $M_{\textup{200m}}$>$4\times10^{14}h^{-1}$\(\textup{M}_\odot\) ATLAS clusters which are also detected in the Abell sample, with no detections in the Planck SZ catalogue.}
	\label{fig:missing_psz_clusters}
\end{figure*}

The lower amplitude of the SZ cluster mass functions compared to $\Lambda$CDM is likely to be due to systematics in the SZ flux-cluster mass scaling relation resulting in an under-estimation of cluster masses of these samples. Indeed, some SZ mass calibration techniques place the Planck cluster counts in closer agreement to the prediction of the $\Lambda$CDM model. However, the incompleteness of SZ samples could be another explanation for the lower amplitude of their mass functions relative to the models. To investigate this possibility, we compare the ACT DR5\footnote{We note that although Planck SZ detections with no measured redshifts are included in the catalogue used in this comparison, at the time of this writing, we only had access to ACT DR5 SZ detections with cluster counterparts identified based on their photometric or spectroscopic redshifts.} and Planck samples to the Abell clusters with richness classes $>2$ and $>3$, (which limit the catalogue to clusters with greater than 80 and 130 members respectively). A summary of our results is presented in Table~\ref{tab:SZ_Abell_comparison}, where we find the SZ samples to recover $\sim30\%$ of Abell clusters with a richness class>2 and $\sim50\%$ of Abell clusters with richness>3. Figure~\ref{fig:missing_psz_clusters} shows examples of four $z\sim0.2$ ATLAS clusters with mass estimates of $M_{\textup{200m}}$>$4\times10^{14}h^{-1}$\(\textup{M}_\odot\) placing them above the Planck lower limit on mass observability. In the future, spectroscopic follow up of SZ clusters guided by optical cluster catalogues could improve the completeness of current SZ samples. Similarly, a  more detailed study of potential issues which could lead to the lack of detection of rich clusters in SZ samples based on comparison to optical catalogues such as ATLAS and redMaPPer would be an interesting topic for future works.

\begin{table*}
	\centering
	\caption[Comparison of ACT DR5 and Planck SZ samples to Abell clusters]{Fraction / percentage of rich Abell clusters overlapping the Planck and ACT DR5 surveys recovered by each SZ survey. ACO2 and ACO3 denote Abell clusters with richness class>2 and 3 respectively.}
	\label{tab:SZ_Abell_comparison}
	\begin{tabular}{c|cc|cc} 
		\hline
		 Redshift & ACT DR5/ACO2 & PSZ/ACO2 & ACT DR5/ACO3 & PSZ/ACO3\\
		\hline
		$0.05<z<0.15$ & 16/73 ($22\%$) & 50/170 ($29\%$) & 4/9 ($44\%$) & 13/29 ($65\%$) \\[0.1cm]
		$0.15<z<0.25$ & 34/105 ($32\%$) & 105/300 ($35\%$) & 13/28 ($46\%$) & 36/64 ($56\%$) \\[0.1cm] 
        \hline
	\end{tabular}
\end{table*}

Although the lower amplitude of SZ cluster mass functions relative to the $\Lambda$CDM predictions could be due to the issues described above, we note that a value of $\sigma_8\approx0.7$ (with $\Omega_\textup{m}=0.3$), could also produce such observations. Furthermore, suppression of observed cluster mass functions compared to $\Lambda$CDM predictions at higher masses (similar to those seen in the ATLAS mass functions) could originate from primordial non-Gaussianities (\citealt{Dalal2008}; \citealt{Verde2010}) or signal the presence of massive neutrinos (e.g. \citealt{Costanzi2013}; \citealt{Castorina2014}; \citealt{Biswas2019}).

While this comparison provides a preliminary indication of the divergence of observations to predictions of the model, a more comprehensive analysis of the systematic and statistical uncertainties on the ATLAS cluster mass function is needed before the statistical significance of the divergence between the observations and the predictions can be quantified. We leave this, as well as comparison of the observations to different mass function models, and obtaining constraints on various cosmological parameters from the ATLAS cluster mass functions to Paper II in this series.

\section{Conclusions}
\label{sec:Conclusions}

In this work, we have presented a new catalogue of photometrically detected galaxy groups and clusters, using the {\sc orca} cluster detection algorithm in combination with the $griz$ bands of the VST ATLAS survey, covering $\sim 4700$ deg$^2$ of the Southern sky. The catalogue contains $\sim22,000$ detections with richness $N_{200}>10$ and $\sim9,000$ clusters with $N_{200}>20$. Using the ANNz2 machine learning algorithm we obtain photometric redshift estimates with an RMS of $\sim0.025$ for our cluster galaxies and a mean redshift uncertainty of $\sim0.01$ for our clusters with richness $N_{200}>20$. The photometric redshift of our sample peaks at $z\sim0.25$, extending up to $z=0.7$. 

We described our calculation of cluster richness ($N_{200}$) which we use as a proxy for cluster mass ($M_{\textup{200m}}$). To this end, we calibrated the mass-richness scaling relation of the ATLAS sample to cluster masses from the SDSS redMaPPer, MCXC, Planck and ACT DR5 samples. We found the ATLAS sample to be $>95\%$ complete at $z<0.35$, $>80\%$ complete up to $z=0.45$, and $\sim60\%$ complete in the $0.45<z<0.65$ redshift range. In terms of cluster mass, we found the sample to be greater than $\sim95\%$ complete for all cluster masses, with near full completeness in the $>5\times10^{14}h^{-1}$\(\textup{M}_\odot\) mass range. Based on a comparison to the SDSS redMaPPer cluster detections as well as visual inspection of our clusters, we estimate the purity of our cluster detections with $M_{\textup{200m}}$>$3\times10^{14}h^{-1}$\(\textup{M}_\odot\) to be $100\%$ in the range $z<0.25$, $87\%$ at $0.25<z<0.35$, $71\%$ at $0.35<z<0.45$ and $56\%$ at $0.45<z<0.55$.

Our comparison to the redMaPPer catalogue showed that in the $0.05<z<0.35$ redshift range, the ATLAS sample contains a larger number of cluster detections and recovers an $\sim40\%$ higher fraction of Abell clusters compared to redMaPPer. At higher redshifts ($0.35<z<0.55$) the redMaPPer catalogue appears to perform better than ATLAS at recovering ACT DR5 clusters, but also detects a large number of clusters not found in the ACT DR5 sample which could be lower mass groups which are misclassified as $\lambda>20$ clusters. At $z>0.35$ we also find a good agreement between the redshift distributions of the ATLAS and redMaPPer samples above a mass limit of $M_{\textup{200m}}$>$3\times10^{14}h^{-1}$\(\textup{M}_\odot\).

We then compared the cluster mass functions of ATLAS, redMaPPer, Planck and ACT DR5 samples to the theoretical predictions of \cite{Tinker2008} models (assuming a \citealt{Planck2016XIII} $\Lambda$CDM cosmology), in the $0.05<z<0.55$ redshift range. At $z<0.35$, ATLAS cluster mass functions have a higher amplitude compared to those of the SZ samples and the redMaPPer mass functions. This places the ATLAS measurements in better agreement with $\Lambda$CDM  predictions with $\sigma_8\approx0.82\pm0.01$ (based on the CMB analysis of \citealt{Planck2016XIII}), rather than some of the previous constraints from the Planck SZ cluster counts, which depending on the SZ mass calibrations, can give $\sigma_8$ measurements as low as $0.71\pm0.03$ \citep{Planck2016XXIV}. Despite this closer agreement, however, at higher masses we found the observed ATLAS mass functions to have a significantly lower amplitude than the $\Lambda$CDM model and the cause of this discrepancy is currently unknown.

Based on our earlier findings, we suggest that the incompleteness of the SDSS redMaPPer sample at lower redshifts could be a contributing factor to the lower amplitude of its mass functions relative to the predictions of $\Lambda$CDM. For the SZ samples, while mass calibration systematics are likely to be the dominant contributing factor to their lower mass function amplitudes relative to $\Lambda$CDM predictions, we show that sample incompleteness is also likely to have a non-negligible contribution and will need to be carefully characterized and accounted for. Future follow-up spectroscopic observations of SZ clusters guided by optical cluster catalogues such as ATLAS and redMaPPer could improve the completeness of SZ samples, while detailed studies of rich optical clusters which remain undetected in SZ samples could help with the identification of any unknown issues impacting the completeness of SZ catalogues. 

In paper II of this series, we shall perform a detailed analysis of the systematic and statistical uncertainties of the ATLAS cluster mass functions. This will in turn enable us to constrain various cosmological parameters including, $\sigma_8$, $\Omega_m$, $\omega$ and $\sum m_{\nu}$ and compare these to constraints from other cluster samples and different cosmological probes.

\section*{Acknowledgements}

BA acknowledges support from the Australian Research Council's Discovery Projects scheme (DP200101068). This work used the DiRAC@Durham facility managed by the Institute for Computational Cosmology on behalf of the STFC DiRAC HPC Facility (www.dirac.ac.uk). The equipment was funded by BEIS capital funding via STFC capital grants ST/K00042X/1, ST/P002293/1, ST/R002371/1 and ST/S002502/1, Durham University and STFC operations grant ST/R000832/1. DiRAC is part of the National e-Infrastructure. For the purpose of open access, the author has applied a Creative Commons Attribution (CC BY) licence to any Author Accepted Manuscript version arising.

The mass function models in Fig.~\ref{fig:SZ_mfs} are generated using the `Halo mass function' module of the {\sc COLOSSUS} python package described by \cite{Diemer2018}. This research has made extensive use of Python 2 \& 3 (\citealt{python2}; \citealt{python3}), IPython \citep{IPython}, Matplotlib \citep{Matplotlib}, SciPy \citep{Scipy}, NumPy \citep{NumPy}, pandas \citep{pandas}, AstroPy \citep{astropy2018}, as well as TOPCAT \& STILTS\footnote{http://www.star.bris.ac.uk/~mbt/topcat/sun253/index.html} packages \citep{Taylor2005}. This research has made use of the NASA/IPAC Extragalactic Database, which is funded by the National Aeronautics and Space Administration and operated by the California Institute of Technology, as well as the "Aladin sky atlas" developed at CDS, Strasbourg Observatory, France.

This publication makes use of data products from the Wide-field Infrared Survey Explorer, which is a joint project of the University of California, Los Angeles, and the Jet Propulsion Laboratory/California Institute of Technology, funded by the National Aeronautics and Space Administration.Funding for the Sloan Digital Sky Survey IV has been provided by the Alfred P. Sloan Foundation, the U.S. Department of Energy Office of Science, and the Participating Institutions. SDSS-IV acknowledges support and resources from the Center for High-Performance Computing at the University of Utah. The SDSS web site is www.sdss.org. The Pan-STARRS1 Surveys (PS1) and the PS1 public science archive have been made possible through contributions by the Institute for Astronomy, the University of Hawaii, the Pan-STARRS Project Office, the Max-Planck Society and its participating institutes, the Max Planck Institute for Astronomy, Heidelberg and the Max Planck Institute for Extraterrestrial Physics, Garching, The Johns Hopkins University, Durham University, the University of Edinburgh, the Queen's University Belfast, the Harvard-Smithsonian Center for Astrophysics, the Las Cumbres Observatory Global Telescope Network Incorporated, the National Central University of Taiwan, the Space Telescope Science Institute, the National Aeronautics and Space Administration under Grant No. NNX08AR22G issued through the Planetary Science Division of the NASA Science Mission Directorate, the National Science Foundation Grant No. AST-1238877, the University of Maryland, Eotvos Lorand University (ELTE), the Los Alamos National Laboratory, and the Gordon and Betty Moore Foundation. This publication makes use of data products from the Wide-field Infrared Survey Explorer, which is a joint project of the University of California, Los Angeles, and the Jet Propulsion Laboratory/California Institute of Technology, funded by the National Aeronautics and Space Administration.

\section*{Data Availability}

The cluster catalogues described in Tables~\ref{tab:cat_centre_cols} and \ref{tab:cat_members_cols} are publicly available at \url{http://astro.dur.ac.uk/cosmology/vstatlas/cluster_catalogue/}. The input galaxy catalogues used for cluster detection are also available from the primary author upon request.  





\bibliographystyle{mnras}
\bibliography{bibliography}



\appendix

\bsp	
\label{lastpage}
\end{document}